\documentclass[5p,sort&compress]{elsarticle}
\usepackage{amssymb,amsmath,xcolor,hyperref,graphicx,multirow,tabularx,physics}
\usepackage{davitex,slashed}
\usepackage{listings,multicol}
\usepackage{anyfontsize}
\usepackage{float}
\usepackage{multicol}
\usepackage{caption}
\usepackage[frozencache=true,section,cachedir=.]{minted}
\usepackage{newfloat}
\DeclareFloatingEnvironment[
  fileext = lol ,
  listname = {List of Listings} ,
  name = Listing
]{code}

\definecolor{bgcolor}{RGB}{247,247,247}
\definecolor{frcolor}{RGB}{227,227,227}
\usepackage[many]{tcolorbox}
\tcbuselibrary{minted}
\newtcbinputlisting{\mycode}[2]{%
  listing engine=minted,
  minted language={#1},
  listing file={#2},
  minted options={
    xleftmargin=3.2em,
    autogobble=true,
    numbers=left,
    numbersep=1em,
    baselinestretch=1.2,
    fontsize=\fontsize{8.5}{8},
    breaklines=true,
    },
  listing only,
  enhanced jigsaw,
  colframe=black,
  sharp corners,
  boxrule=1pt,
  colback=bgcolor, %lightgray!10,
  left=-1.5em,
  left skip=0em,
  width=\textwidth,%-2em,
  overlay={\begin{tcbclipinterior}\fill[frcolor] (frame.south west)
        rectangle ([xshift=5.1mm]frame.north west);\end{tcbclipinterior}},
}

\makeatletter
\newcommand\ff@print[1]{%
  \textnormal{%
    \ifdefined\ff@sf\sffamily\else\ttfamily\fi%
    \ifdefined\ff@bold\fontseries{b}\selectfont\fi%
    #1%
  }%
}
\makeatother

\makeatletter\newcommand\ff@rule
  {\vrule height 6pt depth 1pt width 0pt}
\makeatother

\makeatletter
\ifdefined\ff@noframes\else
  \RequirePackage{tcolorbox}
  \newtcbox\ff@box{nobeforeafter,colframe=gray!80!white,
    colback=bgcolor,boxrule=0.1pt,arc=1pt,
    boxsep=1.2pt,left=0.5pt,right=0.5pt,top=0.2pt,bottom=0.2pt,
    tcbox raise base}
\fi
\makeatother

\makeatletter
\NewDocumentCommand\ff@x{v}{\ff{#1}}
\makeatother

\makeatletter
\newcommand\ff[1]{%
  \ifdefined\ff@noframes%
    \ff@rule\ff@print{#1}%
  \else%
    \relax\ifmmode%
      \ff@rule\ff@print{#1}%
    \else%
      \ff@box{\ff@rule\ff@print{#1}}%
    \fi%
  \fi%
}
\makeatother

\journal{Computer Physics Communications}

\newcommand{\Npf}{N_{\rm pf}}

\begin{document}

\begin{frontmatter}

\title{SIMULATeQCD: A simple multi-GPU lattice code for QCD calculations}

\author[f]{Lukas~Mazur\corref{cor}}
\ead{lukas.mazur@uni-paderborn.de}
\author[b]{Dennis~Bollweg\corref{cor}}
\ead{dbollweg@bnl.gov}
\author[c]{David~A.~Clarke\corref{cor}}
\ead{clarke.davida@gmail.com}
\author[a]{Luis~Altenkort}
\author[a]{Olaf~Kaczmarek\corref{cor}}
\ead{okacz@physik.uni-bielefeld.de}
\author[e]{Rasmus~Larsen}
\author[g]{Hai-Tao~Shu}
\author[d]{Jishnu~Goswami}
\author[b]{Philipp~Scior}
\author[a]{Hauke~Sandmeyer}
\author[a]{Marius~Neumann}
\author[a]{Henrik~Dick}
\author[a,h]{Sajid~Ali}
\author[i]{Jangho~Kim}
\author[a]{Christian~Schmidt}
\author[b]{Peter~Petreczky}
\author[b]{Swagato~Mukherjee\corref{cor}}
\ead{swagato@bnl.gov}

\address[f]{Paderborn Center for Parallel Computing, Paderborn University,
Paderborn, Germany}
\address[b]{Physics Department, Brookhaven National Laboratory,
Upton, New York, United States}
\address[c]{Department of Physics and Astronomy, University of Utah, 
Salt Lake City, Utah, United States}
\address[a]{Fakult\"at f\"ur Physik, Universit\"at Bielefeld,
Bielefeld, Germany}
\address[e]{Department of Mathematics and Physics, University of Stavanger,
Stavanger, Norway}
\address[g]{Institut f\"ur Theoretische Physik, Universit\"at  Regensburg,
Regensburg, Germany}
\address[d]{RIKEN Center for Computational Science, Kobe 650-0047, Japan}
\address[h]{Government College University Lahore, Department of Physics, Lahore 54000, Pakistan}
\address[i]{Institute for Advanced Simulation (IAS-4), Forschungszentrum J\"ulich,
Wilhelm-Johnen-Stra\ss e, 52428 J\"ulich, Germany}
\author{\\[1.5ex] (HotQCD collaboration)}

\begin{abstract}
The rise of exascale supercomputers has fueled competition among GPU vendors, driving lattice QCD developers 
to write code that supports multiple APIs.
Moreover, new developments in algorithms and physics research require frequent updates to existing software. 
These challenges have to be balanced against constantly changing personnel. At the same time, there is a wide
range of applications for HISQ fermions in QCD studies. This situation encourages
the development of software featuring a HISQ action that is flexible, high-performing, open source, easy to use, 
and easy to adapt. In this technical paper, we explain the design strategy, provide implementation details, 
list available algorithms and modules, and show key performance indicators for $\simulat$, a simple 
multi-GPU lattice code for large-scale QCD calculations, mainly developed and used by the HotQCD collaboration. 
The code is publicly available on GitHub.
\end{abstract}


\cortext[cor]{Corresponding authors}
\date{\today}

\end{frontmatter}
% All New Version Announcements must contain the following
% PROGRAM SUMMARY.

\noindent
{\bf Program summary}
  %Delete as appropriate.

\begin{small}
\noindent
{\em Program Title:} SIMULATeQCD                                          \\
{\em CPC Library link to program files:} (to be added by Technical Editor) \\
{\em Developer's repository link:} \url{https://github.com/LatticeQCD/SIMULATeQCD} \\
{\em Code Ocean capsule:} (to be added by Technical Editor)\\
{\em Licensing provisions(please choose one):} MIT  \\
{\em Programming language:} C++, CUDA, HIP                                   \\
{\em Nature of problem:} Quantum chromodynamics (QCD) is the fundamental theory behind the strong force, one of the four fundamental forces in nature. It describes interactions, mediated by gluons, between quarks, the elementary particles that build up protons, neutrons and other hadrons.  One of its unique features is asymptotic freedom: with increasing energy, the interaction strength or coupling between quarks and gluons weakens, which enables studying QCD with peturbation theory. At lower energy scales, however, the strong coupling between quarks and gluons causes perturbation theory to break down, and other tools, such as lattice QCD, need to be employed to obtain QCD predictions.
\\
{\em Solution method:} Lattice QCD is a technique to solve QCD non-perturbatively by discretizing space and time onto a four-dimensional grid. The QCD path integral is thereby rendered finite dimensional and can be evaluated numerically via Monte Carlo methods. In this paper, we present SIMULATeQCD, a C++ code to perform lattice QCD calculations on multiple GPUs, supporting multiple APIs. SIMULATeQCD implements various Monte Carlo sampling methods ranging from sampling of purely gluonic configurations to fully dynamical quark simulations with the Highly Improved Staggered Quark (HISQ) discretization. Modules for measuring various physical observables on these samples are also available.\\
   \\

\end{small}

% ---------------------------------------------------------------------------------------------------- INTRO

\section{Introduction}\label{sec:intro}
Quantum chromodynamics (QCD) is the theory describing the strong force, 
which binds together
quarks to form baryons and mesons.  In the lattice QCD (LQCD) formulation, Euclidean space-time is discretized, 
providing a framework that allows the use of
computational techniques to extract information about physical observables in QCD beyond the weak coupling expansion,
see e.g. Ref. \cite{MILC:2009mpl} for a review.

The lattice formulation of QCD requires the discretization of the QCD action, which leads to discretization effects, e.g. the appearance of additional, unphysical fermion states. It is possible to get rid of these  unphysical states by introducing an additional term in the lattice version
of the Dirac operator as proposed by Wilson~\cite{Wilson:1974sk}. But this discretization
scheme, commonly known as the Wilson fermion action, breaks the chiral symmetry of 
continuum QCD. Another discretization scheme, known as
the staggered fermion formulation~\cite{Kogut:1974ag}, has four tastes
of degenerate mass in the continuum limit. In absence of gauge fields, the four tastes are also degenerate at non-zero lattice spacing. The gauge interaction renders the four tastes no longer degenerate at finite lattice spacing. This is referred to as taste-breaking and is a source of large discretization errors. In practical LQCD calculations
improved staggered actions are used, which are designed to reduce these effects.
Among the improved staggered actions the highly improved staggered quark (HISQ) 
action~\cite{follana_highly_2007}
exhibits the smallest taste-breaking effects \cite{Bazavov:2011nk}. 
The staggered fermion formulation 
preserves a $\U(1)$ subgroup
of the full chiral $\SU(4)_A$ at nonzero lattice spacing, 
Because of this, the quark mass in this formulation
does not have an additive renormalization, unlike in the Wilson formulation. 
This makes the tuning of the bare parameters in this action easier, and because the smallest eigenvalue of
the lattice Dirac operator is bounded from below, this formulation is computationally cheaper,
as the the effort required to produce configurations increases like $m^{-z}$, where $z$ is at least two~\cite{orginos_innovations_2006}.
One recovers a single fermion species per physical flavor by taking
the fourth root of the staggered fermion determinant,
see discussions in Refs.~\cite{Kronfeld:2007ek,Bernard:2007ma}.
An alternative to improved staggered actions is the domain wall
fermion discretization. It avoids unphysical fermion states 
and preserves a lattice version of the chiral symmetry, but it is computationally 
very demanding and is hence not always practical for many lattice calculations.

Since the staggered fermion formulation is computationally the least
expensive approach, has small discretization effects, and preserves a subgroup of the chiral symmetry, it is  the primary choice
for the study of the QCD phase diagram, thermodynamics, and transport properties
of quark-gluon plasma in LQCD, see Refs.~\cite{Ding:2015ona,Schmidt:2017bjt,Guenther:2020jwe} for recent reviews.
In particular, the use of the HISQ action is advantageous in these 
calculations~\cite{Schmidt:2017bjt}.
%The HISQ action has been used by members of the HotQCD collaboration
%to calculate various thermodynamic quantities both at zero and
%non-zero chemical potential, $\mu_B$, including
%the chiral transition temperature
%\cite{Bazavov:2011nk,HotQCD:2018pds},
%the QCD equation of state \cite{HotQCD:2014kol,Bollweg:2022rps,Bollweg:2022fqq},
%the fluctuations of conserved charges \cite{Bollweg:2021vqf,Bazavov:2020bjn,HotQCD:2017qwq,Ding:2015fca,Bazavov:2013uja,Bazavov:2013dta,HotQCD:2012fhj}, the heavy quark potential~\cite{Bala:2021fkm},
%and the screening masses~\cite{Chakraborty:2014aca,Bazavov:2014cta,Petreczky:2019ozv,Bazavov:2019www,Petreczky:2021zmz}.

Highly improved staggered quarks are also
appropriate for situations where one wants to include a dynamical charm quark.
This is relevant for example in determinations of CKM matrix elements, which are
currently being improved with the use of $N_f=2+1+1$ HISQ configurations~\cite{FermilabLattice:2019ikx},
and in recent calculations of the hadronic vacuum polarization contribution
to the muon's anomalous magnetic moment~\cite{FermilabLattice:2022smb}.
Other applications of HISQ action in LQCD calculations include
parton distribution functions, see e.g. Ref. \cite{Fan:2022kcb},
and determinations of the coupling constant $\alpha_s$ and quark masses,
see e.g. Ref.~\cite{Chakraborty:2014aca}.
%\cite{Bazavov:2014soa,Maezawa:2016vgv,Bazavov:2019qoo,Petreczky:2019ozv,Petreczky:2020tky}.
Clearly, there is a wide range of application for LQCD code using HISQ action
to efficiently examine a variety of physically interesting phenomena.

Modern lattice codes tend to be written for GPUs in order to achieve the performance necessary
to make lattice computations feasible. 
In this context it is important that such code is designed to be compatible with multiple APIs, since modern
supercomputers utilize GPUs from multiple manufacturers. In particular, popular Top500 
supercomputers, like Frontier, LUMI, Leonardo, and Perlmutter, utilize AMD or NVIDIA GPUs.
Current lattice calculations also often demand lattice sizes so large that they can
no longer be accommodated in memory by a single GPU, making multi-GPU and multi-node support a fundamental requirement.

One of the most widely used, publicly available production codes
that works on multiple GPUs and features the HISQ action is MILC~\cite{MILC}. 
MILC was originally written
to parallelize using the MIMD paradigm. It is quite reliable
and still used, e.g., in the CKM matrix element and anomalous magnetic moment studies 
mentioned earlier; indeed some of our own design and parameter decisions were modeled
after MILC. On the other hand, owing in part to its long history, it is challenging to
adapt and optimize the native code for new supercomputers, and in fact, MILC achieves
its GPU parallelization by offloading through QUDA~\cite{Clark:2009wm,QUDA}, a lattice
QCD library featuring a wide variety of optimized actions and 
solvers\footnote{Still more lattice codes such as 
Chroma~\cite{Edwards:2004sx,Chroma} and CPS~\cite{CPS}
offload using QUDA and therefore should have access to
the HISQ action. We single out MILC because it is the code base with which
we are most familiar.}.
While the HISQ action already exists in MILC through QUDA,
using lattice-specific external libraries can have its drawbacks.
For example developers need to maintain compatibility with the library,
and there can be a steep learning curve incorporating the library
and understanding its functionalities\footnote{This latter challenge was a reason
we started a new code base in 2017 rather than branching from an existing one.
In the intervening years some lattice code bases have
become significantly more user-friendly.}. 
Moreover specialized applications 
for calculations needed for QCD thermodynamics
are not present or not optimized in MILC. Other popular code bases such as 
Grid~\cite{Boyle:2016lbp,GRID} and openQCD~\cite{openQCD}
do not have the HISQ action implemented and hence cannot be used for our 
purposes.

This motivated us to develop a self-contained, open-source code for the HISQ action
that is designed with the use of multiple GPUs in mind
and particularly suited, but not limited, to study QCD thermodynamics.
Historically, the HotQCD collaboration has primarily used code written by the lattice group of Bielefeld 
University, which has evolved over time from utilizing multiple CPUs to single GPUs. 
Meeting some of the computational challenges described above provided a strong motivation to incorporate multi-GPU support.
Moreover, writing software that meets such computational challenges effectively can be especially difficult
for new developers; hence this need to extend hardware capabilities was taken
as an opportunity to develop a modern, future-proof, multi-GPU framework for LQCD calculations with
completely revised implementations of the basic routines~\cite{Mazur:2021zgi} and structures to
abstract away low-level technical details.
This paper therefore provides implementation details of a new \underline{Si}mple, \underline{Mu}lti-GPU
\underline{Lat}tice code for \underline{QCD} calculations, which we stylize as $\simulat$.
$\simulat$ is available on GitHub \cite{github} and licensed under the MIT license.

$\simulat$ is a publicly available lattice code supporting the HISQ action
that was specifically developed for use on multiple GPUs and for both CUDA and HIP back ends.
It supports $N_f=2+1$ as well $N_f$ degenerate flavors for both pure real and pure imaginary 
baryon chemical potential.
While this code was originally created with HISQ fermions in mind, we would like to stress
that this is not only a HISQ code; indeed it is already able to generate pure SU(3) configurations
and measure a variety of physics observables.
In addition, we have made a special effort to employ a design philosophy that encourages writing highly
modularized code that takes advantage of modern \texttt{C++} features. 
This design strategy helps expedite the implementation of additional high-performing,
parallelized features, such as other action discretizations, going forward.
%The result is easily readable code that has sufficiently 
%abstracted away low-level implementation details, allowing lattice practitioners 
%with intermediate \texttt{C++} knowledge to implement high performing, parallelized methods without much difficulty. 

This paper is organized as follows: \secref{sec:design} explains our design strategy.
Details of the higher-level implementations are given in \secref{sec:modules}, 
followed by the lower-level physics modules in \secref{sec:implement}. 
Finally we showcase the code's performance in \secref{sec:perform} and give a brief outlook in \secref{sec:summary}.

% ---------------------------------------------------------------------------------------------------- DESIGN

\section{Design strategy}\label{sec:design}

\begin{figure}
\centering
\includegraphics[width=\linewidth]{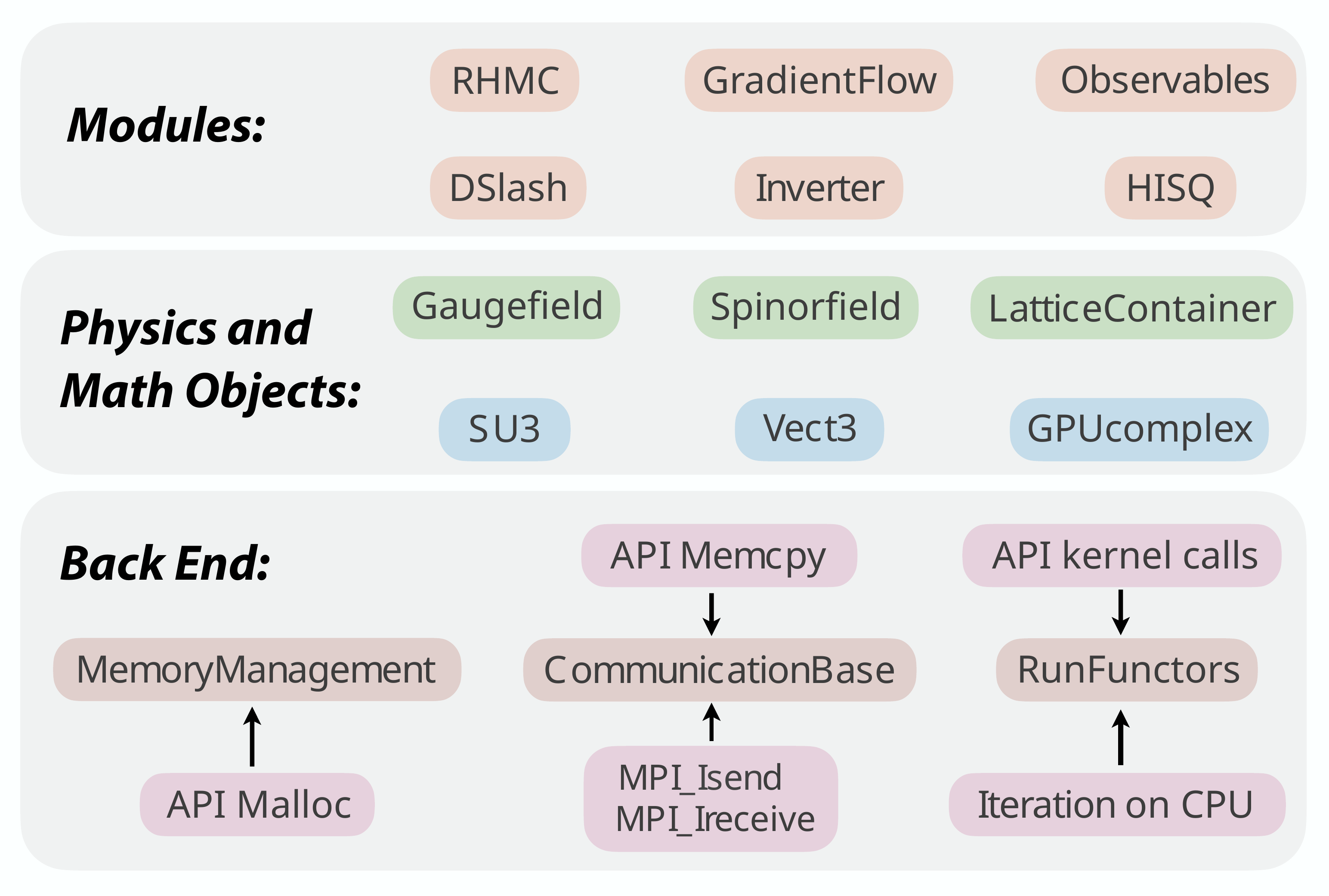}
\caption{Diagram illustrating the code's inheritance hierarchy by listing some example classes. 
Generally speaking, modules inherit from physics and math objects, which in turn inherit from the 
back end. Image adapted from Ref.~\cite{Bollweg:2021cvl}. }
\label{fig:organize}
\end{figure}

%$\simulat$ is a multi-GPU, multi-node lattice code written using \texttt{C++17} and utilizing the
%Object Oriented Paradigm (OOP) and modern \texttt{C++} features. 
In the following we discuss key ideas
guiding the code's design and mention some of the tools already available for
lattice calculations.
In order to adequately address the challenges highlighted in \secref{sec:intro}, we
have worked to develop code that
\begin{enumerate}
  \item is high-performing;
  \item works efficiently on multiple GPUs and nodes;
  \item is flexible to changing architecture and hardware; and
  \item is easy to use for lattice practitioners with intermediate C++ knowledge.
\end{enumerate}
In the remainder of the paper, we refer to these Design Goals as DG1, DG2, DG3, and
DG4, respectively. We try to connect various design choices to these Goals and try
to evaluate our execution of the Goals when we can.

Besides support for multi-GPU via various APIs, the major difference between $\simulat$ and 
its predecessors lies in the clear distinction between organizational levels of the code,
where much of the technical details are hidden from the highest level. This allows physicists without 
advanced C++ or hardware knowledge to write highly
efficient code without having to understand low-level subtleties.
To further help new users, and to improve clarity and reproducibility, we have
extensive documentation on our GitHub. The documentation includes for instance examples how to use
our modules, how to contribute to the code, and internal parameter choices for various algorithms
like the RHMC.

At the highest organizational level are the modules, 
described in \secref{sec:modules}. Here, we strive to write code that closely and obviously
mimics mathematical formulas or short descriptive English sentences. The modules utilize 
general physics and mathematics classes, which sit at an intermediate level. In turn, these classes 
inherit from classes of the back end, which is the lowest organizational level. An overview 
of our code's inheritance scheme is depicted in Fig.~\ref{fig:organize}.

\begin{figure}
\centering
\includegraphics[width=\linewidth]{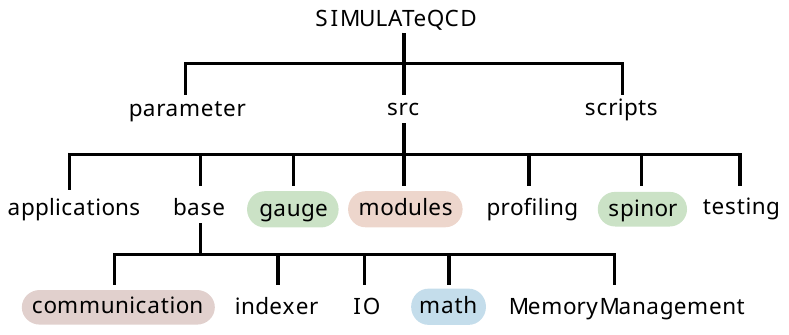}
\caption{Folder layout of $\simulat$.}
\label{fig:folder}
\end{figure}

The folder structure of $\simulat$ is given in Fig.~\ref{fig:folder}.
The root directory features the \texttt{parameter} folder, which contains example files that can
be used as input for executables. We discuss these files in \secref{sec:paramIO}. The \texttt{scripts} folder
contains Bash and Python scripts that assist in using the code, such as
scripts that run tests or estimate the memory usage.
Finally the \texttt{src} folder contains the main source code, which is further partitioned as follows:
\begin{itemize}
    \item \texttt{applications}: Ready-to-use                applications for
          generating configurations and carrying out measurements, see \secref{sec:applications}.
    \item \texttt{base}: Low-level code implementation, which is discussed
          in detail in \secref{sec:implement}.
    \item \texttt{gauge}: Implementation of \ff{Gaugefield} objects,
          described in \secref{sec:implement}.
    \item \texttt{modules}: Modules that are built up from math and physics objects,
          such as the classes for carrying out updates. These are described
          in \secref{sec:modules}.
    \item \texttt{profiling}: Ready-to-use applications to profile and check the performance of 
          $\simulat$, see\\ \secref{sec:testing}.
    \item \texttt{spinor}: Implementation of \ff{Spinorfield} objects,
          described in \secref{sec:implement}.
    \item \texttt{testing}: Ready-to-use applications for testing the code,
          see \secref{sec:testing}.
\end{itemize}

\subsection{Applications}\label{sec:applications}
Due to its design strategy, $\simulat$ makes it easy for developers to 
create new multi-GPU applications out of already existing modules. 
However, many well-tested applications%\footnote{These are not pre-compiled binaries; they need to be compiled.} 
for LQCD calculations already exist. These applications can serve as a starting point for new code, and
they have already been used extensively for production in various physics projects.
Our applications include, but are not limited to, the following:
\begin{itemize}
    \item \texttt{rhmc}: 
    Generate 2+1 flavor HISQ gauge configurations using a rational hybrid Monte Carlo algorithm. 
    Utilized in Refs.~\cite{Clarke:2020htu,Bazavov:2020bjn,Bollweg:2021vqf,Bollweg:2022rps,Bollweg:2022fqq} 
    for $\mu_B=0$ and Refs.~\cite{Dimopoulos:2021vrk,Cuteri:2022vwk} for pure imaginary $\mu_B$. $N_f=3$
    degenerate quarks at $\mu_B=0$ were generated for Ref.~\cite{Dini:2021hug}.
    For details see \secref{sec:confgen}.
    \item \texttt{generateQuenched}:
    Generate pure gauge configurations with the Wilson gauge action by applying heat bath and over-relaxation updates. Utilized in Refs. \cite{altenkort_heavy_2021,sphaleronrate2021,viscosity2022}.
    For details see \secref{sec:confgen}.
    \item \texttt{gradientFlow}:
    Integrate the gradient flow equation using the Wilson or Zeuthen action with fixed or adaptive step sizes 
    and measure various gauge constructs at each step. Utilized in Refs. \cite{altenkort_heavy_2021,sphaleronrate2021,Altenkort:2021jbk,viscosity2022,Altenkort:2023oms}.
    \item \texttt{gaugeFixing}: 
    Fix the gauge of a configuration to the Coulomb or Landau gauge. Can optionally measure various gauge
    constructs at the end. Utilized in Refs.~\cite{Clarke:2019tzf,GaurangParkar:2022aft,GaurangParkar:2022aoa}.
    \item \texttt{wilsonLinesCorrelator}:
    Measure the trace of the product of two Wilson lines going in opposite time directions separated by a distance $r$. Utilized in Refs. \cite{GaurangParkar:2022aft,GaurangParkar:2022aoa}.
    \item \texttt{sublatticeUpdates}:
    Measure the color-electric correlators and energy-momentum-tensor correlators using multilevel algorithm.
\end{itemize}

\subsection{Testing and profiling}\label{sec:testing}
$\simulat$ is constantly being expanded and improved by multiple physicists working in
multiple places of the code, and it is of utmost importance that physics results and performance
remain stable under these changes. To help check and protect against bugs, we have a large
suite of tests as well as some profilers. We only allow code to be merged into the \texttt{main}
branch when it is confirmed that all executables compile and all tests pass. 
Changes to lower level code are asked to further demonstrate no significant loss of performance,
for which we use the profilers.
When a new feature
is implemented, we require the implementer to set up a testing script.
Testing strategies vary depending on the feature, but they include
comparisons to old code; comparisons to analytic results; performing a calculation in
another, independent way; and link-by-link checks against trusted configurations.

% ---------------------------------------------------------------------------------------- PHYSICS MODULES

\section{Physics modules}\label{sec:modules}
A typical lattice calculation is divided into two main parts: First gauge configurations are generated,
then physical observables are measured on the configurations. Many observables are extracted from correlation
functions. Sometimes the observables are extremely noisy, and hence it is necessary to employ 
various noise-reduction techniques. $\simulat$ has modules to accomplish all of these tasks, which we discuss in the
present section.

LQCD uses  a finite region of 
discretized 4-$d$ space-time with spatial extensions $N_\sigma$ and temporal 
extension $N_\tau$, such that the total number of sites is $N_\sigma^3\times N_\tau$. 
Sites are separated by the lattice spacing $a$. 
These quantities are related to the physical volume by $V=(aN_\tau)^3$
and temperature by $T=1/aN_\tau$. 
Each site has coordinates $(x,y,z,t)\in \Z^4$ with $x,y,z \in \lbrace 0, ..., N_\sigma-1\rbrace$ 
and $t\in \lbrace 0, ... , N_\tau-1\rbrace$, and each direction is represented by
$\mu \in \lbrace 0, 1, 2, 3 \rbrace $.
The gluon fields $U_\mu(x)\in\SU(3)$ are the links, and staggered fermion (spinor) 
field objects $\chi^c(x)\in\C^3$~\cite{Kogut:1974ag}, where $c \in \lbrace {0,1,2} \rbrace$,
rest on the sites. Links are implemented 
straightforwardly as $3\times3$ complex matrices 
and the spinor fields as complex 3-vectors. 
These fundamental variables are stored in \ff{Gaugefield}
and \ff{Spinorfield} objects, which contain arrays of links and spinors, respectively. 
\ff{Gaugefield} objects have 
periodic boundary conditions (BCs) while \ff{Spinorfield}
objects have periodic spatial BCs and an anti-periodic time BC.

\subsection{Configuration generation}\label{sec:confgen}

Gauge configurations are generated via Markov Chain Monte Carlo using
Metropolis-type algorithms~\cite{metropolis_equation_1953}. To generate the required
random numbers we use a hybrid Tausworthe 
generator~\cite{press_numerical_1992,lecuyer_maximally_1996}.

Currently, $\simulat$ can generate pure SU(3) configurations and
HISQ\footnote{Our link variable treatment is given in our documentation on GitHub.}
configurations with $N_f=N_l+N_s$ for $N_l$ light and $N_s$ strange dynamical quarks. 
Non-integer numbers of flavors are supported as well.
To enable simulations of four or more degenerate flavors, the fermion determinant was
split up into $\Npf$ pseudofermion fields, since the rational approximation requires $N_i<4$.
\begin{equation}
    \prod^{\Npf}_{i=1} \left( 
    \frac{\det(\slashed{D}_{l,i} + m_l)}{\det(\slashed{D}_{s,i} + m_s)} \right)^{\frac{N_l}{4\Npf}}
    \prod^{\Npf}_{j=1} \left( \det(\slashed{D}_{s,j} + m_s) \right)^{\frac{N_l+N_s}{4\Npf}}
\end{equation}
Note that $N_l$ and $N_s$ are input parameters of the rational approximation, while only $\Npf$
is an input parameter of $\simulat$.
An extension of the code to support different numbers of pseudofermions in the light and strange mass
would be straightforward to implement.
By default, simulations run with $\Npf=1$. Also by default, dynamical quark configurations
are generated at $\mu_B=0$, but it is also
possible to simulate at pure imaginary $\mu_B$~\cite{Hasenfratz:1983ba}.

For pure SU(3) we have implemented the standard Wilson gauge action~\cite{Wilson:1974sk}, with efficient sampling 
through heat bath (HB)~\cite{cabibbo_new_1982,fabricius_heat_1984,kennedy_improved_1985} and 
over-relaxation (OR)~\cite{adler_over-relaxation_1981,creutz_overrelaxation_1987} updates.
For dynamical fermions, we use the HISQ
action~\cite{Follana:2006rc} and generate configurations using
a Rational Hybrid Monte Carlo (RHMC)
algorithm~\cite{Kennedy:1998cu,clark_rhmc_2004}.
To carry out matrix inversions we use a conjugate gradient algorithm, 
for which multiple right-hand sides~\cite{Mukherjee:2014pye} (MRHS), 
multiple shifts \cite{Jegerlehner:1996pm}, and mixed precision implementations are available to improve performance.
The RHMC uses a three-scale integrator~\cite{Sexton:1992nu}, which naturally
profits from the Hasenbusch trick~\cite{Hasenbusch:2001ne}.
By default, the integration uses a leapfrog, but an Omelyan ($2\nd$ order minimal norm)
integrator for the largest scale is also available \cite{OMELYAN2003272}.
\begin{comment}
HISQ fermions utilize two levels of smearing~\cite{blum_improving_1997}.
The first-level link treatment is
\begin{equation}\begin{aligned}
  c_1 &= 1/8\\
  c_3 &= 1/16\\
  c_5 &= 1/64\\
  c_7 &= 1/384,    
\end{aligned}\end{equation}
where $c_1$ is the coefficient for the 1-link, and $c_3$, $c_5$, and $c_7$
are for the 3-link staple, 5-link staple, and 7-link staples, respectively. 
The first-level smeared link is then projected
back to U$(3)$ before the application of the second-level smearing.
The second level uses
\begin{equation}\begin{aligned}
  c_1 &= 1 \\
  c_3 &= 1/16\\
  c_5 &= 1/64\\
  c_7 &= 1/384\\
  c_\text{Lepage} &= -1/8\\
  c_\text{Naik} &= -1/24+\epsilon/8,
\end{aligned}\end{equation}
where $c_\text{Naik}$ and $c_\text{Lepage}$ are the coefficients for the 
Naik\footnote{The code has an explicit $\epsilon$ parameter allowing for an easier, future implementation
of a dynamical charm quark. At the moment it has a default value of zero.}
and Lepage terms~\cite{Lepage:1997id}. This is the same link treatment
used by the MILC collaboration~\cite{MILC:2010pul}.
We use the HISQ/tree action, which is a tree-level improved 
L\"uscher-Weisz action in the gauge sector~\cite{MILC:2009mpl}. The relative
weights of the plaquette and rectangle terms are 
\begin{equation}\begin{aligned}
    c_\text{plaq} &= 5/3, \\
    c_\text{rect} &= -1/12.
\end{aligned}\end{equation}
\end{comment}
The value of the parameter in the force cut-off that we use in the force filter is $\delta = 10^{-5}$~\cite{MILC:2010pul}. This value may need to be tuned when one wants to perform QCD thermodynamics calculations very close to the chiral limit.
\subsection{Inverters}
For matrix inversions, as needed for example in the RHMC algorithm, we implement multiple types of the Conjugate Gradient (CG) algorithm. 
The CG inverter solves the linear equation $Ax=b$ where $A$ is a symmetric, positive definite matrix, $x$ is the solution vector and $b$ the input vector. In our practical applications, $A$ is almost always the $M^{\dagger}M$ operator where $M$ is the fermion matrix. To separate our inverter implementation of this specific case however,  our implementations contain \ff{invert} methods that take an abstract base class \ff{LinearOperator}, as well as two vector field templates representing $x$ and $b$. The \ff{LinearOperator} class' only member function is the virtual \ff{applyMdaggM} function that  defines the interface for the matrix vector product $Ay$ that will be used in the inversion. The two main classes that are used in our production setting are \ff{ConjugateGradient} and \ff{AdvancedMultiShiftCG} with the former offering a standard as well as multi right-hand-side version of the CG algorithm which solves the linear equation $Ax_{k}=b_{k}$ with multiple input vectors $b_k$. The \ff{AdvancedMultiShiftCG} class provides the multi-shift CG algorithm that we use in our RHMC algorithm. It solves the linear equation $\left(A+\sigma_{k}\right)x_{k}=b$ for various shifts $\sigma_{k}$. It exploits the shift invariance of the Krylov subspace constructed in the conjugate gradient iteration so that the expensive matrix vector product, here the \ff{applyMdaggM} call, only needs to be performed on the smallest shift $\sigma_0$ and solutions for higher shifts $\sigma_i,\;\;i>0$, are obtained along the way at the cost of few additional AXPY operations \cite{Jegerlehner:1996pm}.

\subsection{Measurement of observables}
Implementations for physical observables consisting of gauge constructs include: the Polyakov loop,
Polyakov loop correlators in the singlet, average, and octet channels, the chiral condensate,
the topological charge, topological charge density, its correlator, and the Weinberg 
operator
calculated from the field strength tensor and $\order{a^2}$- and $\order{a^4}$-improved field strength
tensor~\cite{bilson-thompson_highly_2003}, 
the plaquette,
the clover,
the color-electric and color-magnetic correlator, and
energy-momentum tensor correlators.

%There is a method that measures the chiral condensate for a small number of random vectors.
%This method is used to report {\it in situ} measurements of the chiral condensate while
%the RHMC is running.
Hadronic correlation functions along any direction are implemented using HISQ fermions (point sources) 
to compute, for example, (screening) masses of mesons in various channels.
We also support measurement of Taylor coefficients for the expansion of $\log\ZQCD$ in $\mu_B/T$.

Some observables, such as certain Polyakov loop correlators, must be measured in special gauges.
To this end, Coulomb and Landau gauge fixing are implemented
via over-relaxation~\cite{mandula_efficient_1990} updating,
with over-relaxation parameter $\omega=1.3$.

\subsection{Noise reduction}
For applications where one needs to attenuate short-distance gauge fluctuations,
we have implemented the gradient flow~\cite{luscher_properties_2010} with Wilson and 
Symanzik-improved actions~\cite{ramos_symanzik_2016} (Zeuthen flow). The numerical integration is
carried out using a third-order Runge-Kutta method 
for Lie groups~\cite{munthe-kaas_runge-kutta_1998,celledoni_commutator-free_2003} with fixed or adaptive step sizes.
The blocking method~\cite{Altenkort:2021jbk} can be a good supplement to the gradient flow method
to substantially improve the signal-to-noise ratio of bosonic correlators or other correlators with
disconnected contributions. It is implemented by breaking two planes at different time slices into
bins and computing bin-bin correlators. 

We have also implemented hypercubic blocking (HYP) smearing \cite{Hasenfratz:2001hp}, 
which uses hypercubic fat links to minimize the largest fluctuations of the plaquette 
by maximizing the smallest plaquettes. 

The multi-level algorithm~\cite{luscher_locality_2001,meyer_locality_2003} is implemented by incorporating 
sublattice updates and observable calculation. The lattice in the temporal direction is divided uniformly 
into sublattices. Within each sublattice, HB and OR updates are performed in 
parallel several times followed by a measuring of the observable. At the moment the Polyakov loop, 
color-electric correlators, and energy-momentum tensor correlators can be measured using this algorithm.

\subsection{General calculation of all-to-all correlators}

One of the most common types of observable needed to calculate in lattice computations
is the correlator. Generically a correlator is an operator
\begin{equation}
    \text{correlator}=\ev{f(A,B;r)}\Big|_D,
\end{equation}
where $A$ and $B$ are some operators evaluated at space-time positions $x$ and $y$
belonging to the domain $D$, $r=|x-y|$, and $f$ is an arbitrary function of $A$ and $B$.
In order to calculate this quantity, one has to find all possible
pairs of sites at a given $r$. To save the user from having to solve this problem
for their own correlator of interest, in pursuit of DG4,
we have implemented a general framework.

The \ff{CorrelatorTools} class handles these general calculations. The fundamental
method is the \ff{correlateAt} method
\begin{equation}
    \texttt{correlateAt<f>("domain",A,B)}.
\end{equation}
This correlates arrays \ff{A} and \ff{B} of arbitrary type
according to the {\it correlator archetype} \ff{f}. The site pairing scheme
is controlled by \ff{domain}. This returns an array of $\ev{f(A,B)}$
indexed by $r^2$. At the moment, this framework is only available for single GPU use.

\section{Low-level implementation}\label{sec:implement}
The back end of \simulat~operates close to the hardware level to achieve high 
performance (DG1), for example through dedicated classes for memory management and communication between host and devices, as lattice applications are usually bound by bandwidth. 
The four-dimensional lattice also needs to be translated to one-dimensional memory, which is handled by an indexing class.
Furthermore, there is a class that maps sets of lattice sites to GPU threads, such that operations on multiple sites are run in parallel. The big advantage of this class is that the operations on each site can be written as high-level functors that do not require any knowledge about the specifics of GPU programming. The back end of the code is finally completed with classes that manage file input/output and logging.
 
\subsection{Allocating and deallocating memory}

The memory demands for lattice calculations can increase with increasing lattice size
in a nontrivial way, depending on the algorithm being used. Thus for complex applications, it can be difficult to keep track of all dynamically allocated memory on both host and device, especially since GPU memory allocation and deallocation has to be handled via the appropriate API, depending on the hardware. Additionally, to avoid memory
leaks, memory should be automatically deallocated when appropriate. Finally, memory (de)allocation can take a non-negligible amount of time,
so one can gain a performance boost by allowing large chunks of memory to be shared.

Addressing these issues in a way that is straightforward for lattice
practitioners without detailed knowledge of HPC touches on all four Design Goals.
Therefore, we developed the centralized \ff{MemoryManagement} class,
which manages and knows about all instances of dynamically allocated memory. This is done through the
\ff{gMemory} member class, which can hold the raw pointer to dynamically allocated memory along 
with API-independent wrappers for memory allocation and deallocation. \ff{gMemory}
objects are pointed to by our custom smart pointer, the \ff{gMemoryPtr}. 
A \ff{gMemoryPtr} is labelled by a \texttt{SmartName}, a descriptive string chosen by the user that
by default prevents that two pointers point to the same memory.
However, the memory can be shared by choosing a \texttt{SmartName} beginning with the string \texttt{SHARED}.
Each \ff{gMemoryPtr} object also informs the \ff{MemoryManagement} when it is created and destroyed. 

The \ff{MemoryManagement} is never explicitly instantiated, as all necessary methods are static. 
This includes those needed to create new dynamic memory, destroy it, and
report to the user which dynamic memory exists, along with its size and
\texttt{SmartName}.

\subsection{Communication}

\begin{figure}[t!]
\centering
\includegraphics[width=\linewidth]{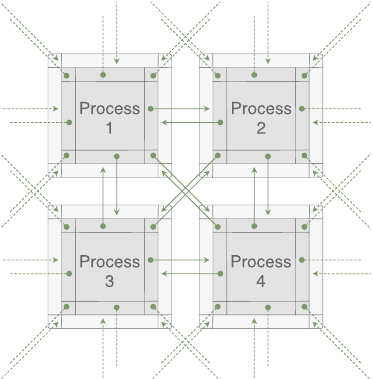}\\
\vspace{10mm}
\includegraphics[width=\linewidth]{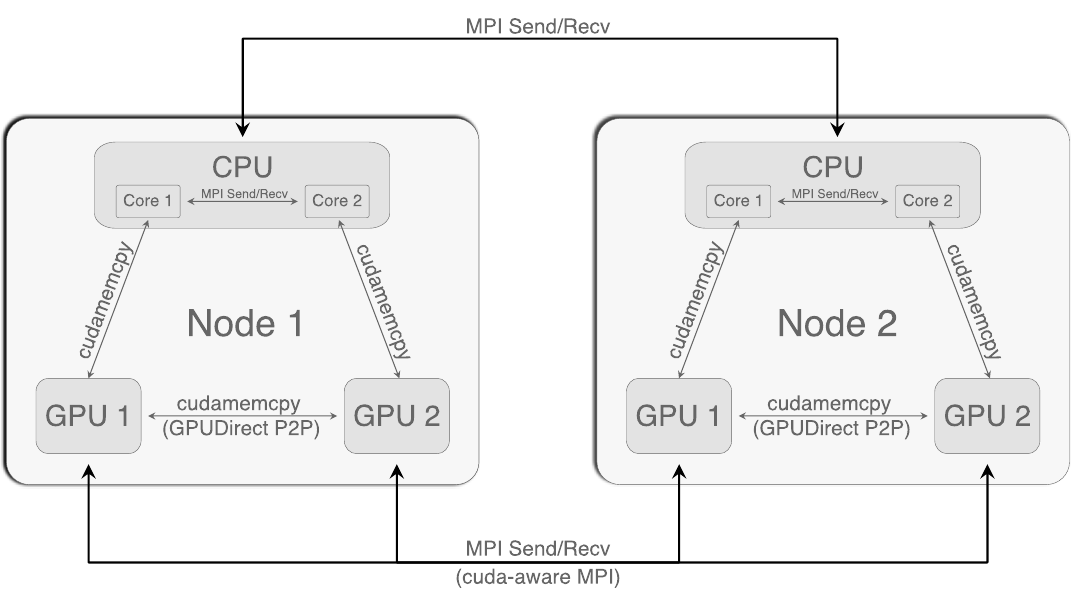}
\caption{{\it Top}: Schematic halo exchange for four processes in two dimensions, each process containing
         a sublattice. The bulk is indicated by the dark gray squares, and the halo is indicated by
         light grey squares. Copies of sites located near the 
         borders of the bulk, set off here by solid grey lines, are stored in the halos.
         Information starts at solid green circles and is injected into the halos following the arrows.
         Dotted lines indicate injections following periodic boundary conditions. {\it Bottom}: Illustration of 
         the topology of a fictitious HPC system which consists of two nodes with two (NVIDIA) GPUs per node. 
         The arrows represent all possible communication paths. Images taken from Ref.~\cite{Mazur:2021zgi}.}
\label{fig:communicate}
\end{figure}

To address DG2, $\simulat$ splits a lattice into multiple
sublattices, with partitioning possible along any
direction. Each sublattice is given to a single GPU, and we
call this sublattice the {\it bulk}.
In addition, the GPU holds a copy of the outermost borders of neighboring sublattices, which we call the {\it halo}. This halo is necessary because many
measurement and update processes are stencil operations, which means that a
calculation performed at the boundary of the sublattice may need information from a neighboring
sublattice. Hence, boundary information from all sublattices must be copied
into their neighbors' halos. A schematic drawing of the exchange 
of halo information between different sublattices is shown in 
Fig.~\ref{fig:communicate} (top).

\begin{figure}
\centering
\includegraphics[width=\linewidth]{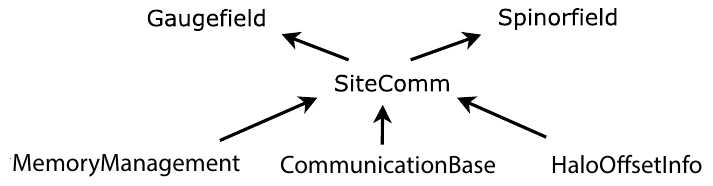}
\caption{Dependence structure of the \ff{Gaugefield} and \ff{Spinorfield} on communication classes. }
\label{fig:siteComm}
\end{figure}

\begin{code*}[!htb]
\mycode{c++}{functor.cpp}
\caption{An example functor, \ff{PlaquetteKernel}, utilizing functor syntax. It is templated to allow for arbitrary precision \ff{floatT}, to run on GPU with \ff{onDevice==True}, for arbitrary \ff{HaloDepth}, and to allow the \ff{Gaugefield} to use arbitrary \ff{CompressionType}. This functor takes a \ff{Gaugefield} object as argument, whose elements are accessed in memory through the \ff{gaugeAcessor} \ff{gAcc}, set to point to the \ff{Gaugefield} accessor in the initializer list. The argument of \ff{operator()} indicates that this functor will be iterated over \ff{gSite} objects. We indicate with \ff{All} that we run over both even and odd parity sites. The method \ff{getLinkPath} multiplies all links starting at \ff{site}, following a path in the specified directions. We compute the real part of the trace with \ff{tr\_d}. Due to our functor syntax, all lattice splitting, communication, and distribution to GPU threads is done behind the scenes.}\label{lst:functor}
\end{code*}

\begin{code*}[!htb]
\mycode{c++}{useIterator.cpp}
\caption{Using an iterator to calculate the plaquette with the functor given in Listing~\ref{lst:functor}. 
We instantiate our \ff{Gaugefield} object and a \ff{LatticeContainer}, which is needed for the reduction across threads and nodes. Our iterator, \ff{iterateOverBulk} will assign each \ff{gSite} in the bulk of a sublattice to a GPU thread; hence it needs to have $N_{\sigma,\tinysp\text{sub}}^3\times N_{\tau,\tinysp\text{sub}}=\texttt{vol4}$ elements. The reduction happens over the global lattice, so in the last time we normalize by $N_\sigma^3\times N_\tau$, three colors, and six plaquettes per site in four dimensions.}\label{lst:iterator}
\end{code*}

\begin{code*}[!htb]
\mycode{c++}{functorMagicComplete.cpp}
\caption{Sketch of low level functor and iterator implementation. API-dependent constructions are wrapped inside \ff{performFunctor} and \ff{iterateFunctor} methods. The user can decide at compile time whether to use the CUDA or HIP API, which is implemented here using \ff{ifdef}. These methods are templated to allow memory access to an arbitrary return type \ff{res} using memory accessor \ff{Accessor}. The operation \ff{op} is at this level completely unspecified; it will be defined at a higher level through a functor, for example the plaquette functor shown in Listing~\ref{lst:functor}. The types \ff{CalcReadInd} and \ff{CalcWriteInd} respectively translate from GPU thread index to input memory index and from input to output memory index. The \ff{site} object in \ff{performFunctor} is set to \ff{auto} to allow use with arbitrary indexing object, for example the \ff{gSite} and \ff{gSiteMu} objects defined in \secref{sec:indexing}.}\label{lst:magic}
\end{code*}

\begin{code*}[!htb]
\mycode{c++}{generalOperator.cpp}
\caption{An excerpt from the \ff{GeneralOperator} implementation
     showing the specialization of \ff{GeneralOperator} to add two ``Class" objects.}\label{lst:GeneralOperator}
\end{code*}

Communication between CPU processes is facilitated through MPI, which also enables GPU communication. For both NVIDIA and AMD GPUs, intranode communication is managed using P2P \ff{memcpy} calls, provided by the respective vendor libraries (CUDA and HIP). For internode communication, we utilize GPU-aware MPI when available on the cluster (commonly referred to as CUDA-aware MPI on NVIDIA HPC clusters). On clusters where GPU-aware MPI is not available, we resort to using a normal version of MPI, and hence, GPU-GPU internode communication occurs through the CPUs. 
An illustration of different communication paths for two nodes is provided in Fig.~\ref{fig:communicate} (bottom).
Since halos extend the boundary area around a sublattice, they may use a significant amount of memory. Additionally, the P2P and GPU-aware MPI implementations require additional communication buffers on the GPU compared to the GPU-unaware MPI implementation. Therefore, if communication is not a limiting factor for a specific application, it might be beneficial to turn off GPU-aware MPI/P2P to save GPU memory and use the GPU-unaware MPI communication instead. 
Hence the code also provides options to choose which communication path should 
be used (DG1).

Halo communication proceeds by first copying halo information contiguously into a buffer.
This requires translating from the sublattice's indexing scheme to the buffer's indexing
scheme. The \ff{HaloOffsetInfo} class provides offsets for different halo segments 
(stripe halo, corner halo, etc). These offsets and the buffer 
base pointer are used to place the halo data at the correct position in the buffer. 
An example for the corner halo would be:
\begin{equation}\begin{aligned}
\text{corner buffer pointer} = ~&\text{buffer base pointer}\\ 
                                   &+ \text{corner halo offset}. \nonumber
\end{aligned}\end{equation}

The \ff{SiteComm} class is the lowest-level class from which all objects that need to communicate across sublattices, such as the \ff{Gaugefield} introduced in the next section,
inherit. It uses the \ff{MemoryManagement} class to allocate memory for the 
buffer\footnote{In this context there are a few different kind of send/receive buffers for the halo, depending on the
communication scheme, e.g. GPU-aware MPI or P2P. To help manage this we also 
have a \ff{HaloSegmentInfo} class, which adds the halo offset to the pointer for the corresponding buffer.},
uses the \ff{HaloOffsetInfo} to translate the local index to the buffer index, copies information into the buffer,
and finally uses the \ff{CommunicationBase} to carry out the exchange\footnote{More precisely, we use the \ff{CommunicationBase} in case
MPI or GPU-aware MPI is used for communication. In the case of P2P, we just call CUDA/HIP \ff{memcpy} 
to carry out the communication.}. This chain of dependencies is illustrated in Fig~\ref{fig:siteComm}.

Additionally, we enhance performance (DG1) by allowing the code to perform certain computations concurrently with communication, such as copying halo buffers into the bulk, whenever feasible.

\subsection{Indexing and fields}\label{sec:indexing}

Our site indexing is done in lexicographic order,
but for many purposes, it is convenient to characterize sites with an even/odd parity\footnote{Sometimes 
referred to as a checkerboard with red/black sites.}. 
For applications where this is the case, it is more efficient to organize sites in memory 
such that all even sites are in the first half of the memory and all odd sites are in the second half. 
Therefore we convert the 4-$d$ coordinate of a site to the 1-$d$ memory index as
\begin{equation}\begin{aligned}
    \text{index}=&~\frac{1}{2}\left(x+yN_\sigma+zN_\sigma^2+tN_\sigma^3\right)\\
    &+\frac{1}{2}N_\sigma^3N_\tau\left(x+y+z+t\right)\bmod2.
\end{aligned}\end{equation}
Building up from this, links are indexed by
\begin{equation}\begin{aligned}
    \text{index}=&~\frac{1}{2}\left(x+yN_\sigma+zN_\sigma^2+tN_\sigma^3\right)\\
    &+\frac{1}{2}N_\sigma^3N_\tau\left(x+y+z+t\right)\bmod2 + \mu N_\sigma^3N_\tau.
\end{aligned}\end{equation}

When using multiple GPUs, similar formulae hold for the sublattices, except that the
extensions are replaced by $N_\sigma\to N_{\sigma,\tinysp\text{sub}}$ and
$N_\tau\to N_{\tau,\tinysp\text{sub}}$. Hence we distinguish between the local index and global index.
Moreover, as explained in the previous section and as depicted in Fig.~\ref{fig:communicate}, information
about neighboring fields is stored in a halo surrounding the bulk. Therefore besides bulk indexing,
each sub-lattice has so-called ``full'' indices that include the halos.
This scheme is computed by
\begin{equation}\begin{aligned}
         \text{index}_{\texttt{\tinysp Full}}= &~ \frac{1}{2}\left(x+y N_{x}+z N_{x}N_{y} +t N_{x}N_{y}N_{z}\right) \\
          &+\frac{1}{2}N_{x}N_{y}N_{z}N_{t}\left(x+y+z+t\right)\bmod2,
\end{aligned}\end{equation}
while the links are indexed by
\begin{equation}\begin{aligned}
         \text{index}_{\texttt{\tinysp Full}}= &~\frac{1}{2}\left(x+y N_{x}+z N_{x}N_{y} +t N_{x}N_{y}N_{z}\right)\\
          &+\frac{1}{2}N_{x}N_{y}N_{z}N_{t}(x+y+z+t)\bmod2\\ 
          &+ \mu  N_{x}N_{y}N_{z}N_{t},
\end{aligned}\end{equation}
with
\begin{equation*}
     \begin{aligned}
         N_{i}= & ~N_{\sigma}+H_{i},\qquad i\in x,y,z,\\
         N_{t}= & ~N_{\tau}+H_{t},
    \end{aligned}
\end{equation*}
where $H_i$, $H_t$ are the halo depths in different directions.

In order to abstract these difficulties away from the user, we have implemented \ff{gSite}
objects. A \ff{gSite} object holds all information about a site, including its coordinates
and index in the local, global, and full context.
All methods for calculating indices, coordinates, and movement along the lattice are contained
in the \ff{GIndexer} class. With DG4 in mind, we make it easy for the user
to control indexing by passing a
\ff{Layout} template parameter, which can be set to \ff{Even}, \ff{Odd}, or \ff{All}.
Similarly, the \ff{gSiteMu} class holds all information about link indexing.

\subsection{Functor syntax}

Much of the effort in this code goes into abstracting away highly complex parallelization, which depends 
on the API, whether GPUs or CPUs are used, the number of processes, the node layout being used, and so on.
Accomplishing this for the general case is crucial to DG4; hence we have 
implemented a system where one can iterate an arbitrary operation that depends on 
arbitrary arguments over an arbitrary set of coordinates, taking care of the
parallelization behind the scenes.

One common task in lattice calculations is to perform the same calculation on each lattice site, 
which involves link variables at or close to that site, for example, computing the plaquette, 
\begin{equation}
     \plaq_{\mu\nu}(x)=U_\mu(x)U_\nu(x+a\hat{\mu})
                        U^\dagger_\mu(x+a\hat{\nu})U^\dagger_\nu(x).
\end{equation}
An example {\it kernel} computing the plaquette is shown in \autoref{lst:functor}.
The code for the kernel is wrapped in a {\it functor}, i.e. a \ff{struct} 
with its \ff{operator()} overloaded, which in this case takes a \ff{gSite} object as argument. 
The functor is passed to a function which we call an {\it iterator}. 
This function iterates over all required lattice sites and calls the functor on each one, thereby distributing the calls on the available computing 
resources. Basic iterators are provided by the \ff{RunFunctors} class which implements iteration over the bulk, the full lattice including halos, as well as a looping iterator that can iterate over the lattice and an additional index. Physics classes like the \ff{Gaugefield} and \ff{Spinorfield} that derive from \ff{RunFunctors} use these to create specialized iterators to iterate for example over all links of a \ff{Gaugefield} in a specific direction or a subset of RHS for our multi-RHS \ff{Spinorfield}.  An example usage of this is given in Listing~\ref{lst:iterator}, where the \ff{iterateOverBulk} iterator is called using the functor \ff{plaquetteKernel} as a template parameter.
By using functors as template parameters for iterators, we can conveniently iterate arbitrary 
calculations over sites, without the need to write specialized code each time. 

Iterators such as \ff{iterateOverBulk}, which calls the
\ff{struct}'s \ff{operator()} at bulk sites, inherit from the class
\ff{RunFunctors}, which contains the lowest-level methods that iterate
functors over the desired target set. In Listing~\ref{lst:magic} we provide sketches of the most salient methods
in the \ff{RunFunctors} class, namely \ff{performFunctor}, which applies a
functor to a \ff{gSite} or \ff{gSiteMu} object, and \ff{iterateFunctor},
which carries out \ff{performFunctor} using CUDA or HIP. 

If we are not using GPUs and just evaluate the computation kernel on multiple CPUs, then the iterator method 
iterates sequentially over the sites/links of a sublattice using \ff{for}-loops, with one sublattice 
per CPU core in parallel. But if we use GPUs, then there are no \ff{for}-loops. Instead, the computation of 
each iteration of these nested \ff{for}-loops will also be parallelized on threads of the GPUs. 
For example with the plaquette computation,
\begin{equation}
    {\rm number\ of\ GPU\ threads} = \frac{N_\sigma^3\times N_\tau}{{\rm number\ of\ GPUs}}
\end{equation}
are spawned, where each thread is doing the work corresponding to a certain site.

As mentioned in \secref{sec:indexing}, we index sites and links through \ff{gSite} and \ff{gSiteMu} 
objects, respectively; hence one usually passes either a \ff{gSite} or \ff{gSiteMu} object
to the functor. The \ff{runFunctors} class also contains several
\ff{CalcGSite} methods that tell the iterator how to translate from
these objects to GPU thread indices. 

\subsection{Expression templates}
To facilitate an efficient but intuitive composition of math expressions involving basic physics objects such as spinorfields and
gaugefields (DG4), we implement general math operations such as additions, subtractions,
multiplications and divisions via expression templates \cite{Veldhuizen:1995}.

We define a set of functions such as \ff{general\_add} and \ff{general\_mult} that, instead of performing the 
corresponding math operation, return a functor \ff{GeneralOperator} that holds 
the information to carry out the desired calculation. The execution of this calculation is triggered by the 
copy assignment operator (=) of the basic physics object, which passes the functor to an \ff{iterateFunctor} method.
 
Using the ``Substitution failure is not an error" (SFINAE) principle, we have specialized
the \ff{GeneralOperator} struct for different type combinations such as ``Class + Class" and ``Class + scalar". In this
context, ``Class" means an object that has a \ff{getAccessor}-method and operator() defined. This object could be 
a \ff{Gaugefield}, a \ff{Spinorfield} or another \ff{GeneralOperator}.
Nesting of expressions such as $a\cdot b+c\cdot d$, where $a,b,c,d$ are physics objects, and combinations of operators
and scalars such as $(a\cdot b)\cdot2.0$ are also implemented. This highly templated, 
low-level code allows users 
to write high-level physics code involving \ff{Spinorfield}s, \ff{Gaugefield}s, etc. 
that closely resembles
familiar mathematical equations while also avoiding superfluous evaluations of temporary expressions.

Listing~\ref{lst:GeneralOperator} shows an excerpt of the \ff{GeneralOperator} implementation detailing the specialization
of additions of two class objects.

\subsection{LatticeContainer and reduction}\label{sec:LatCont}

In lattice QCD computations, intermediate results per lattice site are typically calculated. These intermediate results are then reduced over the entire lattice, such as by summing over the entire lattice. When a lattice is divided among multiple processes, the procedure becomes slightly more complex:

\begin{enumerate}
   \item Intermediate results are computed in parallel on all sites of all sub-lattices.
   \item Each process performs a reduction into a single quantity.
   \item A final reduction is performed across all processes into a single quantity, which is then distributed over all processes if necessary.
\end{enumerate}

In order to address DG4, this common task has been combined into a single reduction method which is part of the \ff{LatticeContainer} class. This class is a container of dimension $N_{\sigma}^3\times N_\tau$ which elements can be of any datatype and is distributed over the processes similarly to the\\ \ff{Gaugefield} or \ff{Spinorfield}. By calling that reduction method, these elements are automatically reduced across all GPUs.

In addition to a standard sum over all elements, the \ff{LatticeContainer} class also provides a method to reduce only the time slices of the lattice, resulting in a vector of length $N_\tau$. This is particularly useful for computing correlation functions in the time direction.
Furthermore, there is the stacked reduction, which can be used to reduce the intermediate results of a stacked spinor. 
Internally, all these reduction operations are performed using MPI's collective communication calls in combination with CUB (NVIDIA) and hipCUB (AMD).

\subsection{Parameter handling and IO}\label{sec:paramIO}

Parameter files hold input parameters such as quark masses and couplings, as well as information
like the lattice dimension, random number seed, etc. These are implemented through
our \ff{LatticeParameters} class. More specialized parameter classes,
for example the parameter class for the RHMC, inherit from this.
Default parameter files are in the \texttt{parameter} folder
mentioned in \secref{sec:design}, but one can also pass a custom
parameter file as argument to any executable.
\begin{comment}
Another challenge facing large-scale LQCD projects is finding the storage
space for petabytes of configurations. Furthermore, configurations are highly flexible
to many kinds of measurements, so it is conducive for open science to store
these configurations in a standardized, accessible way, find long-term storage elements
on which to place them, and to responsibly collect and curate metadata to help
lattice practitioners find these configurations, for example through a search tool.
In this context, it is nowadays stressed that scientific data should be
FAIR~\cite{wilkinson_fair_2016}.

The most well equipped scheme to effectively carry out such a task in LQCD
is the 
International Lattice Data Grid (ILDG)~\cite{beckett_building_2011},
which provides a common format and metadata scheme
in order to facilitate configuration sharing among lattice groups,
and which is now being resuscitated~\cite{Karsch:2022tqw}.
ILDG configuration binaries must be written out in LIME
format~\cite{LIME}
\end{comment}

$\simulat$ supports the International Lattice Data Grid (ILDG) format for reading and writing the gauge configurations~\cite{beckett_building_2011}. 
The gauge configuration
files for ILDG format are packaged together with with a corresponding XML
metadata (QCDml) file~\cite{maynard_qcdml_2005}.
To assist with forming a QCDml file, one can pass optional, ILDG-compliant
parameters to the \ff{LatticeParameters} class that are printed
to the standard output. These metadata can then be captured by custom scripts.
In addition to ILDG format, $\simulat$ also has support for MILC~\cite{MILC}, NERSC,
and openQCD~\cite{openQCD}
configuration formats, three other commonly used formats in the lattice community.

%\begin{figure}[t!]
%    \centering
%    \includegraphics[width=\linewidth]{MRHS_DSLASH_JUWELS_DATA.pdf}
%    \caption{Performance of the HISQ Dirac operator with varying number of RHS vectors on a 
%    single JUWELS Booster node.}
%    \label{fig:multirhs}
%\end{figure}

\begin{figure}[t!]
    \centering
    \includegraphics[width=\linewidth]{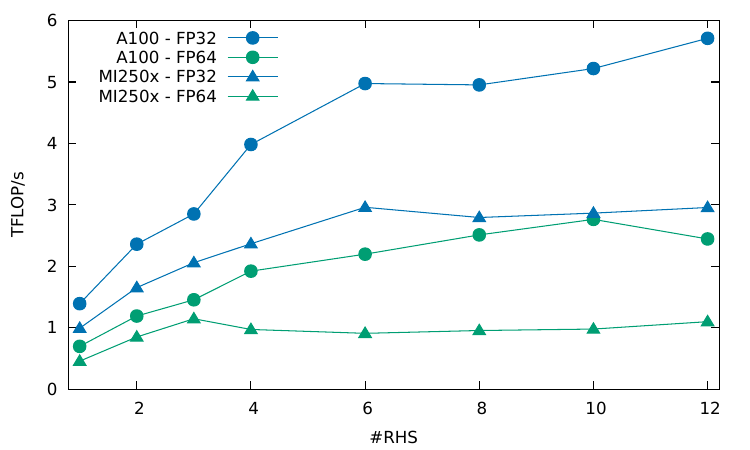}
    \caption{Performance of the HISQ Dirac operator with varying number of RHS vectors on a $32^4$ lattice on a single A100 GPU and single MI250x GCD.}
    \label{fig:multirhs}
\end{figure}

\begin{figure}[t!]
    \centering
    \includegraphics[width=1\linewidth]{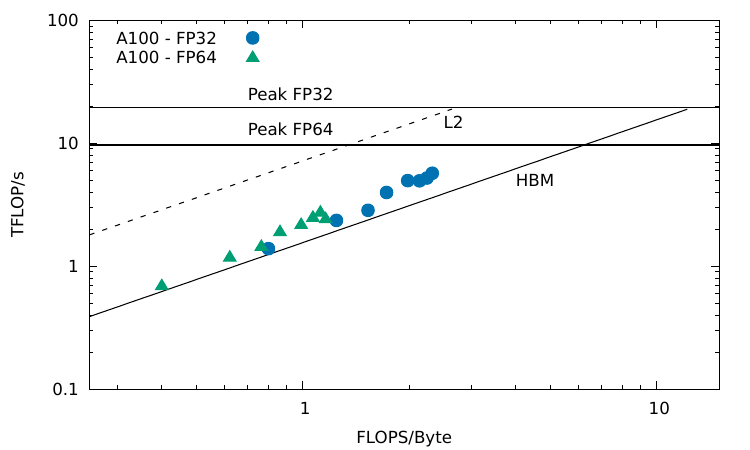}
    \includegraphics[width=1\linewidth]{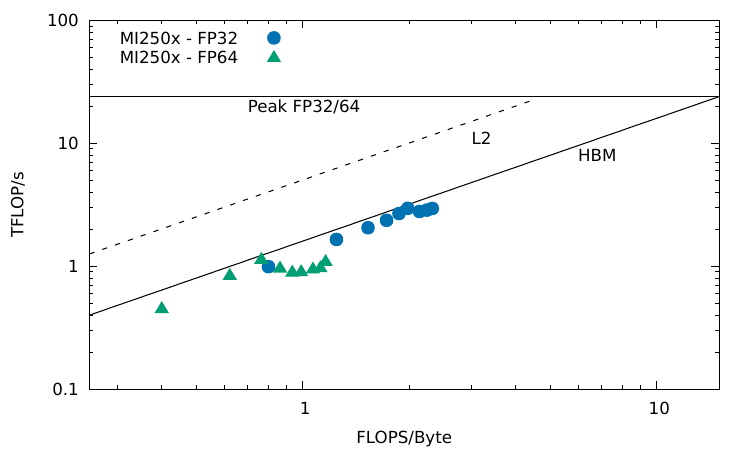}
    \caption{Roofline plots of the HISQ Dirac operator with varying number of RHS vectors on a $32^4$ lattice on a single A100 GPU and single MI250x GCD.}
    \label{fig:roofline}
\end{figure}

\begin{figure}[t!]
\centering
\includegraphics[width=\linewidth]{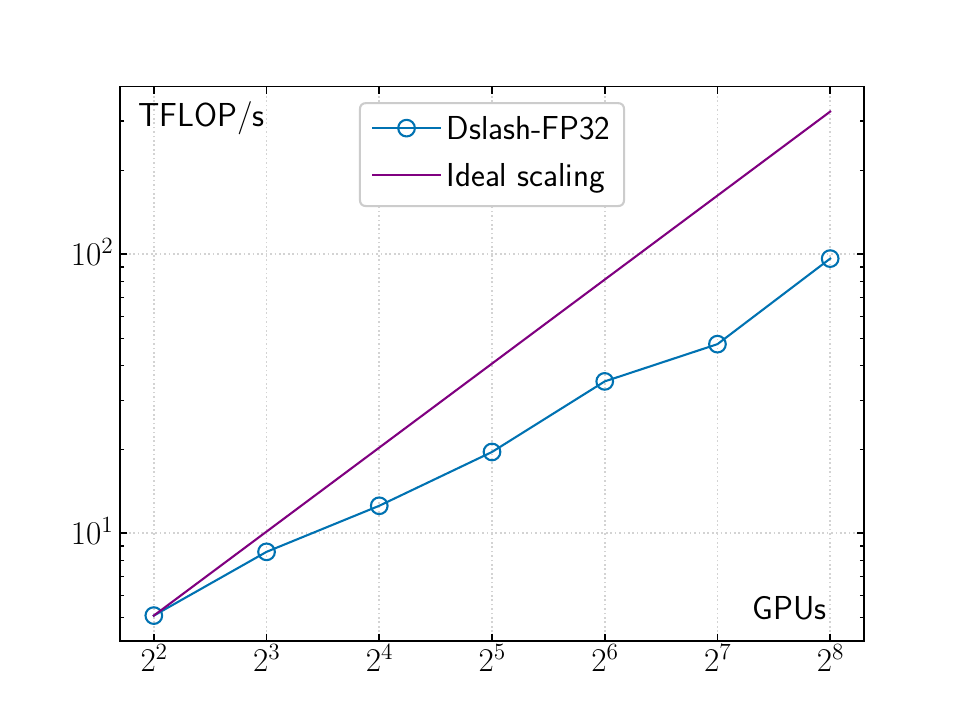}\\
\includegraphics[width=\linewidth]{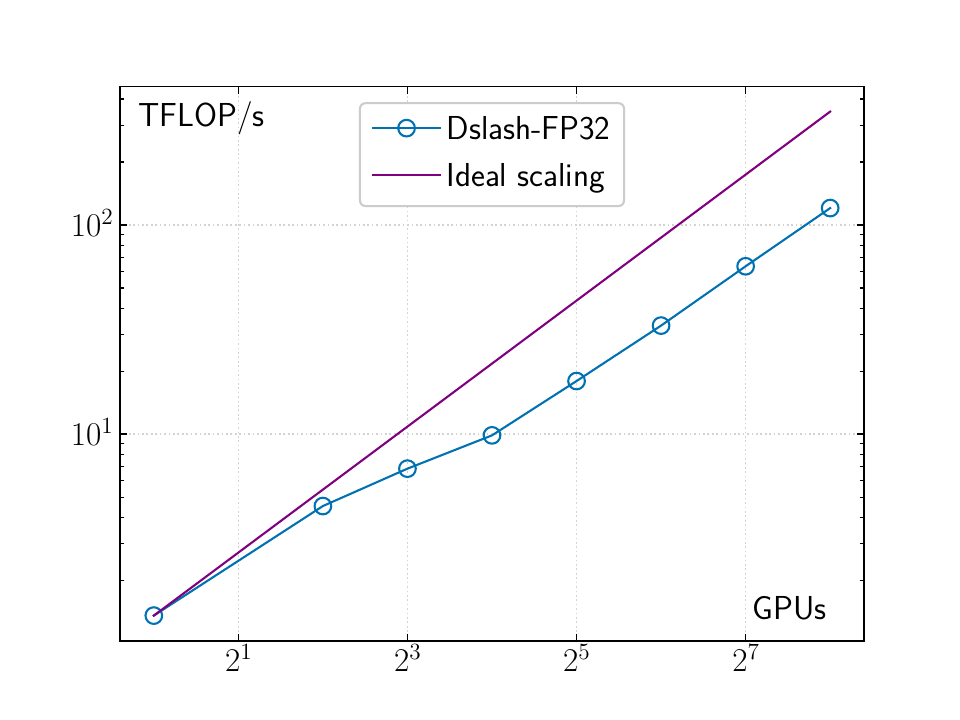}
\caption{Scaling of HISQ Dirac operator with a single RHS on Perlmutter. {\it Top}: 
Strong scaling for a $96^4$ lattice. {\it Bottom}: Weak scaling for a $32^4$ local lattice.}
\label{fig:scaling_pm}
\end{figure}

\begin{figure}[t!]
\centering
\includegraphics[width=\linewidth]{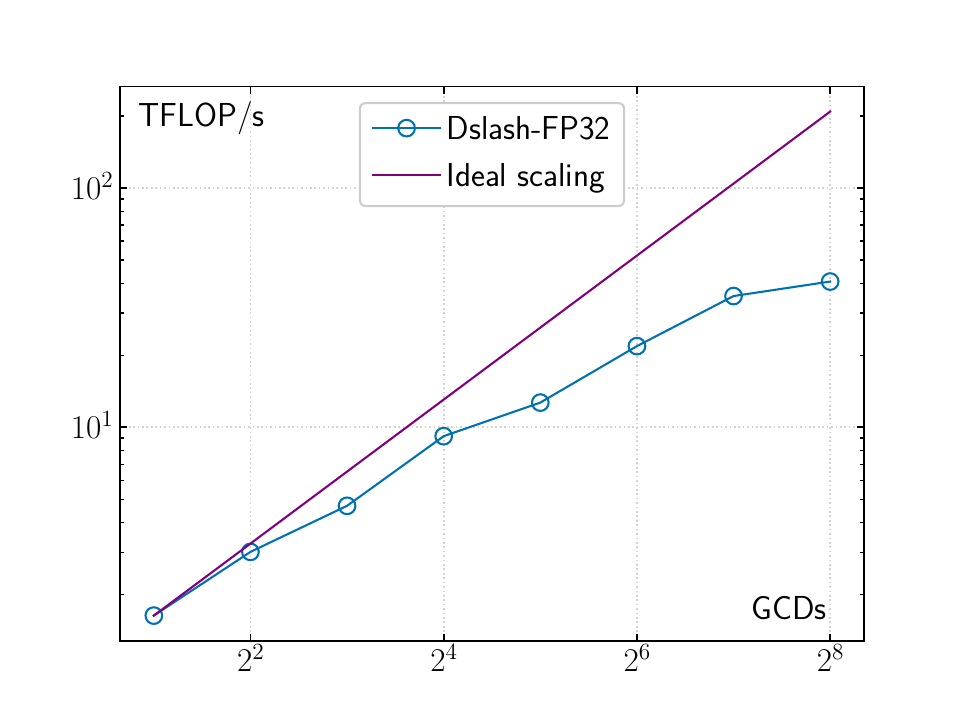}\\
\includegraphics[width=\linewidth]{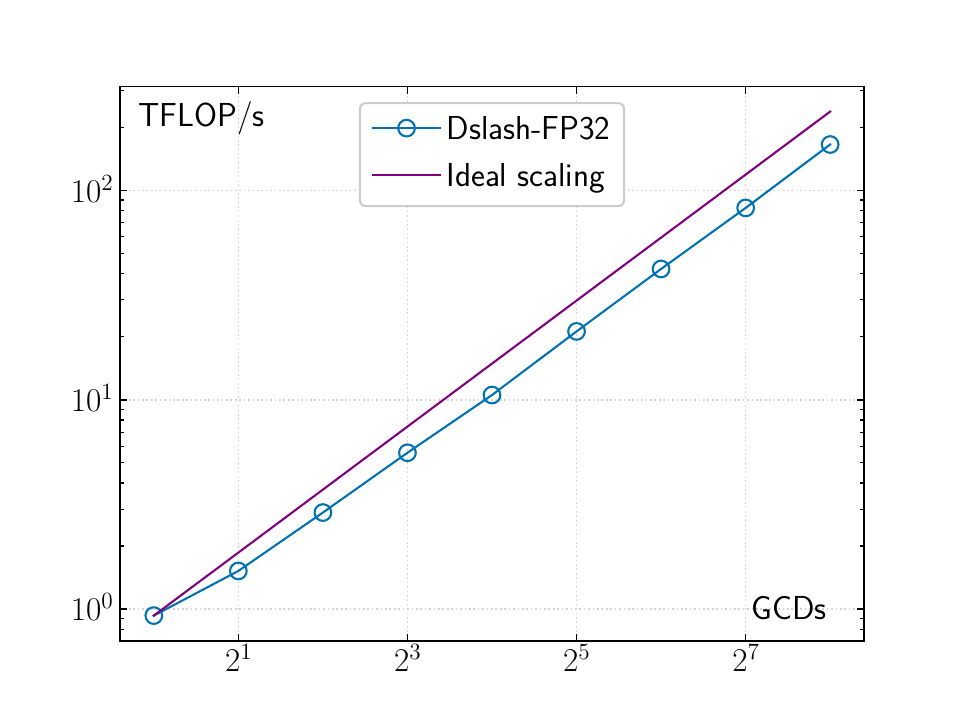}
\caption{Scaling of HISQ Dirac operator with single RHS on Frontier. {\it Top}:
Strong scaling for a $96^4$ lattice. {\it Bottom}: Weak scaling for a $32^4$ local lattice.}
\label{fig:scaling_frontier}
\end{figure}

% -------------------------------------------------------------------------------------- PERFORMANCE
\section{Performance}\label{sec:perform}

In this section we showcase our code's performance, giving quantitative
insight on our ability to achieve DG1.

%In order to save memory without losing too much precision, especially in performance-critical
%situations, we use link compression.
%For instance our HISQ smearing implementation 
%needs several temporary \ff{Gaugefield} objects, and for the Naik
%term it is not necessary to represent a link with 18 reals.
%Uncompressed gauge fields can be reconstructed for example through unitarization.

When generating HISQ configurations for typical parameters, about 60\% of our RHMC run time is spent
inverting the Dirac matrix via conjugate gradient, and in it, applying the $\slashed{D}$ operator 
to a vector is the most performance-critical kernel. Hence this and related kernels are important benchmarks. 
%The kernel's performance is limited by the available memory bandwidth, but its arithmetic intensity 
%can be increased by applying the gauge field to multiple RHS simultaneously\footnote{The MRHS
%inverter is not used for the RHMC. It is more useful for the calculation of observables such
%as generalized susceptibilities.}. 
%Furthermore, it benefits from gauge field compression. 
%Only a subset of the link matrix entries are stored in memory, and the 
%missing entries are recomputed from the stored ones based on the symmetries, either $\SU(3)$ or $\U(3)$, 
%of the link matrix.

%To improve performance, we overlap certain computations, such as the kernels that prepare halo buffers,
%with communication.

\subsection{HISQ Dslash performance model}
To understand the HISQ Dslash's performance, we will use a simple roofline model. The action of the kernel on a vector $\phi_x$ at a given lattice site $x$ takes the form
\begin{align*}
    \label{eq:dslashstencil}
    \slashed{D}\phi_x&=\sum_{\mu=0}^{4}\bigg[\left(X_{x,\mu}\phi_{x+\hat{\mu}}-X^{\dagger}_{x-\hat{\mu},\mu}\phi_{x-\hat{\mu}}\right)\\
    &+\left(W_{x+3\hat{\mu}}\phi_{x+3\hat{\mu}}-W^{\dagger}_{x-3\hat{\mu},\mu}\phi_{x-3\hat{\mu}}\right)\bigg],
\end{align*}
where $X$ and $W$ are smeared gaugefields and we ignore constants and the staggered phases for convenience. In each direction, the kernel performs four complex three dimensional matrix-vector multiplications as well as three vector additions which results in 1146 floating point operations (FLOP) per site x. Meanwhile, the kernel has to read 288 floats for the link matrices $X$ and $W$,  96 floats for the vectors at $x\pm\hat{\mu}$ and $x\pm3\hat{\mu}$ and write 6 floats at site $x$. This results in an arithmetic intensity of 
\begin{align*}
    \mathrm{FLOP/BYTE}(\slashed{D})&=\frac{1146\;\mathrm{FLOP/site}}{(288+96+6)\mathrm{words/site}},   
\end{align*}
which translates to arithmetic intensities of about 0.73 FLOP/Byte in single precision and 0.36 FLOP/Byte in double precision. As the ratio of theoretical peak FLOP/s to memory bandwidth of modern GPUs is around $\mathcal{O}(10)$ FLOP/Byte, the HISQ Dslash's performance is heavily bound by the devices memory bandwidth. This can be improved in situations where the kernel can be applied to multiple right-hand-side (RHS) vectors at once. The loaded gaugefields $X$ and $W$ can then be re-used, resulting in an arithmetic intensity that scales with the number of RHS vectors like
\begin{align*}
    \mathrm{FLOP/BYTE}(\slashed{D})_{n-RHS}&=\frac{n\cdot1146\;\mathrm{FLOP/site}}{(288+n\cdot(96+6))\mathrm{words/site}}.   
\end{align*}
Our implementation offers two methods to accommodate applying the Dslash kernel to multiple RHS vectors: an inner loop over $N_{\mathrm{inner}}$ RHS vectors inside each thread of the kernel so that data re-use via registers or L1 cache is possible as well as an outer loop over $N_{\mathrm{outer}}$ RHS vectors as an additional thread-block dimension in which case re-use can only happen via the L2 cache. Both can be combined so that the kernel is applied to $N_{\mathrm{inner}}*N_{\mathrm{outer}}$ RHS vectors.
With the above arithmetic intensity at hand, we can benchmark our HISQ Dslash implementation and compare the achieved performance against the performance rooflines stemming from the GPUs peak FLOP/s as well as memory bandwidths, which we show in the following two subsections.
\subsection{Benchmarks on A100 GPUs}
The performance of our multi-RHS Dslash implementation for varying number of right hand side vectors on a single A100 GPU is shown in Fig.~\ref{fig:multirhs}. For each number of RHS vectors, we tested all possible combinations of $N_{\mathrm{inner}}$ and $N_{\mathrm{outer}}$ and show the best performing combination in the plot.

We see significant performance improvements from\\ Gaugefield re-use with increasing number of RHS vectors up to 6 RHS vectors and a further slight increase in performance up to 12 RHS vectors in the single precision case. For double precision, we see a more gradual increase in performance up to 10 RHS vectors. For both precision, we found the best performance when keeping $N_{\mathrm{inner}}$ and $N_{\mathrm{outer}}$ of similar size but avoid $N_{\mathrm{inner}}>6$ as that leads to register spilling.

In Fig.~\ref{fig:roofline} we compare our kernels performance to\\ rooflines dictated by the A100's theoretical peak FLOP/s as well as HBM DRAM and L2 cache bandwidths. We find that for both FP32 and PF64 arithmetic and all number of RHS vectors, we surpass the performance ceiling imposed by the A100's HBM bandwidth of 1.55 TB/s. Closer inspection of the kernels performance using NVIDIA's Nsight compute profiler shows that we saturate about 85\% of the the theoretical peak HBM bandwidth but efficient cache usage lifts our kernels effective bandwidth above that limit. 
 
 % The scaling on a single node is close to ideal and we achieve a maximum of about 11.4 TFLOP/s on a full node while using
%8 RHS.  

%Profiling the Dslash kernel using NVIDIA's Nsight compute software reveals that we achieve
%a memory throughput of up to 1.36 TB/s on a single A100 GPU, thus coming very close to the
%cards peak memory bandwidth of about 1.55 TB/s. 

In Fig.~\ref{fig:scaling_pm} we show strong- and weak-scaling benchmarks of our HISQ DSlash on Perlmutter,
which has 4 NVIDIA A100 GPUs per node. Nodes are interconnected via HPE Slingshot 11 fabric and have
4 NICs providing 25GB/s bandwidth each. We find good weak scaling on this system
up to 256 GPUs. A slight decrease in speedup can be seen as soon as node-to-node communication
starts. 
Strong scaling benchmarks with a $96^4$ global lattice size start deviating from ideal scaling
earlier. As the local lattice sizes and thus the compute workloads of the individual GPUs keep getting smaller,
hiding the HISQ DSlash's communication becomes increasingly difficult.

\subsection{Preliminary Benchmarks on MI250X GPUs}
The performance of our Dslash implementation on a single MI250x GCD for varying number of RHS vectors is shown in Fig.~\ref{fig:multirhs} with the corresponding roofline plot in the bottom half of Fig.~\ref{fig:roofline}. For single precision arithmetic, we can see an increase in performance up to 6 RHS vectors after which the performance plateaus, whereas for double precision arithmetic, performance increases only up to 3 RHS vectors and then flattens. In contrast to our A100 benchmarks, we only observe a significant performance gain from using multiple RHS when we iterate over the vectors inside the inner loop, i.e. keeping $N_{\mathrm{outer}}=1$. We suspect that this is due to the MI250x's much smaller L2 cache of 16MB compared to the A100's 40MB L2 cache. In the roofline plot, we can see that we stay consistently below the HBM bandwidth imposed performance limit. 
In Fig.~\ref{fig:scaling_frontier} we show HISQ DSlash benchmarks on Frontier. This system is configured with 4 
AMD MI250X per node. Each MI250X is comprised of 2 Graphic Compute Dies (GCDs); i.e. there are 8 GCDs per node. 
GCDs inside an MI250X are connected via Infinity Fabric with 200GB/s bi-directional bandwidth, and the 4 
MI250X cards on a node are connected via Infinity Fabric with bi-directional bandwidths between 50-100GB/s 
depending on their arrangement.
Nodes are connected via four HPE Slingshot NICs each providing 25GB/s of bandwidth. 
The weak scaling benchmark with a $32^4$ local lattice volume shows close to ideal scaling all the way 
up to 256 GCDs. Going from 8 GCDs to 16 GCDs shows no
significant drop in speedup although node-to-node communications kick in.
The strong scaling benchmark scales well up until 16 GCDs. After that point, communication can no longer 
be hidden effectively and the speedup decreases. 
First tests on LUMI-G produced comparable single GCD performance as that in Frontier.
%We note that the single GPU performance achieved on MI250X is currently lagging behind what we would 
%expect given the cards specifications. A single GCD has
%a memory bandwidth similar to that of a single A100 GPU, and we would therefore expect to see performance
%differences between both to be smaller. As profiling tools and ROCm compilers for AMD GPUs mature, we are investigating what is 
%causing this decreased performance and continue to optimize our code for this new hardware.

%\begin{table}[]
%    \centering
%    \begin{tabular}{c|c|c|c|c}
%    no. of RHS & 4 GPUs &  3 GPUs &   2 GPUs &  1 GPU \\\hline
%1 & 4.73079 & 3.64066 & 2.43864 & 1.27963 \\
%2 & 7.66433 & 5.93165 & 3.95775 & 2.10292\\
%3 & 9.3739 & 7.30559 & 4.89197 & 2.66587\\
%4 & 10.5115 & 8.20125 & 5.54996 & 3.06388\\
%5 & 11.0659 & 8.64023 & 5.87168 & 3.27174\\
%6 & 11.317 & 8.82256 & 6.06521 & 3.3924\\
%7 & 11.0128 & 8.6586 & 5.98085 & 3.37701\\
%8 & 11.4201 & 9.00431 & 6.27061 & 3.57343\\
%9 & 10.7651 & 8.54542 & 5.94003 & 3.36035\\
%10 & 11.1947 & 8.86835 & 6.18 & 3.51152\\
%    \end{tabular}
%    \caption{Data for Figure~\ref{fig:multirhs}, showing the performance of the HISQ Dirac operator on a JUWELS Booster node for different %numbers of right hand side vectors (RHS).}
%    \label{tab:mrhs}
%\end{table}

\begin{table}[]
\centering
\begin{tabular}{c|c|c|c|c}
 & \multicolumn{2}{c}{A100} & \multicolumn{2}{c}{MI250x}\\
RHS & FP32 & FP64 & FP32 &FP64\\\hline
1 & 1.36 & 0.70 & 0.93 & 0.45\\
2 & 2.36 & 1.19 & 1.65 & 0.85\\
3 & 2.85 & 1.45 & 2.06 & 1.14\\
4 & 3.98 & 1.92 & 2.36 & 0.97\\
6 & 4.97 & 2.20 & 2.96 & 0.91\\
8 & 4.95 & 2.51 & 2.79 & 0.95\\
10 & 5.22 & 2.76 & 2.86 & 0.98\\
12 & 5.71 & 2.44 & 2.95 & 1.10\\
\end{tabular}
\caption{Data for Figure~\ref{fig:multirhs}, showing the performance in TFLOP/s of the HISQ Dirac operator with varying number of RHS vectors on a $32^4$ lattice on a single A100 GPU and a single MI250x GCD. }
\end{table}

%\begin{table}
%\centering
%    \begin{tabular}{c|c||c|c}
%    \multicolumn{2}{c}{strong scaling} & \multicolumn{2}{c}{weak scaling} \\
%GPUs  & TFLOP/s & GPUs & TFLOP/s \\\hline
%1 & - & 1 & 1.35746 \\
%4 & 5.06892 & 4 & 4.53759 \\
%8 & 8.58352 & 8 & 6.85441  \\
%16 & 12.5533 & 16 & 9.87794 \\
%32 & 19.5635 & 32 & 17.9565 \\
%64 & 35.0211 & 64 & 33.0596 \\
%128 & 47.6174 & 128 & 63.4786\\
%256 & 96.4686 & 256 & 120.445
%    \end{tabular}
%    \caption{Data for Figure ~\ref{fig:scaling_pm}, showing the strong and weak scaling of the HISQ Dirac operator on Perlmutter with a %$96^4$ global lattice and $32^4$ local lattice, respectively.}
%    \label{tab:pmscaling}
%\end{table}

\begin{table}
\centering
    \begin{tabular}{c|c||c|c}
    \multicolumn{2}{c}{strong scaling} & \multicolumn{2}{c}{weak scaling} \\
GPUs  & TFLOP/s & GPUs & TFLOP/s \\\hline
1 & - & 1 & 1.36 \\
4 & 5.07 & 4 & 4.54 \\
8 & 8.58 & 8 & 6.85  \\
16 & 12.55 & 16 & 9.88 \\
32 & 19.56 & 32 & 17.96 \\
64 & 35.02 & 64 & 33.06 \\
128 & 47.62 & 128 & 63.48\\
256 & 96.47 & 256 & 120.44
    \end{tabular}
    \caption{Data for Figure ~\ref{fig:scaling_pm}, showing the strong and weak scaling of the HISQ Dirac operator on Perlmutter with a $96^4$ global lattice and $32^4$ local lattice, respectively.}
    \label{tab:pmscaling}
\end{table}

%\begin{table}[]
 %   \centering
 %   \begin{tabular}{c|c||c|c}
%\multicolumn{2}{c}{strong scaling} & \multicolumn{2}{c}{weak scaling} \\
%GCDs  &  TFLOP/s  &  GCDs  &  TFLOP/s\\\hline
%1  &  -  &  1  &  0.92974\\
%2  &  1.63049  &  2  &  1.51439\\
%4  &  3.00675  &  4  &  2.89454\\
%8  &  4.69513  &  8  &  5.57654\\
%16  &  9.17246  &  16  &  10.5269\\
%32  &  12.6743  &  32  &  21.2049\\
%64  &  21.8216  &  64  &  42.1544\\
%128  &  35.3419  &  128  &  82.5013\\
%256  &  40.6282  &  256  &  165.718\\
%    \end{tabular}
%    \caption{Data for Figure ~\ref{fig:scaling_frontier}, showing the strong and week scaling of the HISQ Dirac operator on Frontier with a %$96^4$ global lattice and $32^4$ local lattice, respectively.}
%    \label{tab:frontierscaling}
%\end{table}
\begin{table}[]
    \centering
    \begin{tabular}{c|c||c|c}
\multicolumn{2}{c}{strong scaling} & \multicolumn{2}{c}{weak scaling} \\
GCDs  &  TFLOP/s  &  GCDs  &  TFLOP/s\\\hline
1  &  -  &  1  &  0.93\\
2  &  1.63  &  2  &  1.52\\
4  &  3.01  &  4  &  2.89\\
8  &  4.69  &  8  &  5.58\\
16  &  9.17  &  16  &  10.53\\
32  &  12.67  &  32  &  21.20\\
64  &  21.82  &  64  &  42.15\\
128  &  35.34  &  128  &  82.50\\
256  &  40.63  &  256  &  165.72\\
    \end{tabular}
    \caption{Data for Figure ~\ref{fig:scaling_frontier}, showing the strong and week scaling of the HISQ Dirac operator on Frontier with a $96^4$ global lattice and $32^4$ local lattice, respectively.}
    \label{tab:frontierscaling}
\end{table}

% ------------------------------------------------------------------------------------- CONCLUSIONS

\section{Outlook}\label{sec:summary}

We presented \simulat,  a multi-GPU, multi-node lattice code that allows simulation of dynamical 
fermions using the HISQ action and works for both NVIDIA and AMD back ends.
We chose the HISQ action since it is widely used for state-of-the-art, high-statistics QCD studies.
In writing \simulat, we sought to write clean code that is highly modularized, so that anyone with modest
knowledge of C++ can get started writing production code right away. We believe this is an important and 
sometimes overlooked characteristic of code in a lattice context, where personnel and hardware are
constantly changing.
At the same time, we did our best to ensure good performance by writing code close to the 
hardware that is both mindful of and flexible to the computing system.
We leveraged functor syntax to balance both of these needs.
\simulat\ has many production-ready applications, which are listed in \secref{sec:applications}. We gave
some details on the implementation of our physics modules in \secref{sec:modules} for easy and transparent 
reference in the future.

What exists in \simulat\ presently mostly includes
code that has been relevant to HotQCD projects.
One of the most important characteristics of $\simulat$ is that it is relatively
straightforward to implement new algorithms; hence wherever there is desire and manpower,
our code can be extended to suit the needs of a new project in an
uncomplicated, undemanding way.
One of the obvious avenues of extension for $\simulat$ on the physics side includes implementation of other 4-$d$ 
actions\footnote{Domain wall fermions will require an overhaul of the existing indexer, or
the implementation of a new one.},
such as the Wilson action for fermions.
While the code already works for $N_f=2+1$ and degenerate $N_f$ systems,
some work is still needed to include a dynamical charm quark. 
A relatively new feature allows measurement of Taylor coefficients of the pressure,
however this still needs porting a deflated solver algorithm to be suitable for efficient use with lighter quarks.
Finally, the code is already relatively integrated with ILDG,
and this should ideally be continued in parallel with the ILDG revival.

Turning to low-level implementation, machines such as Aurora are utilizing Intel GPUs, so it will be
valuable to introduce also a Sycl back end. Along this vein, while most modules 
require a GPU back end, everything can also be compiled and run on CPUs 
only\footnote{It should be noted that at the moment, we have not yet implemented vectorized CPU code.} 
by adding a few macros, which so far has been done for a few selected modules. 

\section*{CRediT authorship contribution statement}

\textbf{Lukas Mazur:} Conceptualization, Methodology,\\ Project administration, Software, Validation. Writing -- Review \& Editing.
\textbf{Dennis Bollweg:} Software, Validation. Writing -- Review \& Editing.
\textbf{David A. Clarke:} Software, Validation. Writing -- Original Draft, Review \& Editing.
\textbf{Luis Altenkort:} Software, Validation. Writing -- Review \& Editing.
\textbf{Olaf Kaczmarek:} Conceptualization, Supervision, Project administration, Funding acquisition.
\textbf{Rasmus Larsen:} Software. 
\textbf{Hai-Tao Shu:} Software. 
\textbf{Jishnu Goswami:} Software.
\textbf{Philipp Scior:} Software, Validation. 
\textbf{Hauke Sandmeyer:} Conceptualization, Software. 
\textbf{Marius Neumann:} Software. 
\textbf{Henrik Dick:} Software. 
\textbf{Sajid Ali:} Software. 
\textbf{Jangho Kim:} Software.
\textbf{Christian Schmidt:} Software, Supervision, Funding acquisition.
\textbf{Peter Petreczky:} Writing -- Review  \&  Editing.
\textbf{Swagato Mukherjee:} Supervision,\\ Funding acquisition.

\section*{Acknowledgements}

This material is based upon work supported by the U.S. Department of Energy, Office of Science, 
Office of Nuclear Physics through Contract No. DE-SC0012704, and within the framework of 
Scientific Discovery through Advance Computing (SciDAC) award Fundamental Nuclear Physics 
at the Exascale and Beyond. 

This research used resources of the Oak Ridge Leadership Computing Facility, which is a DOE Office of Science User Facility supported under Contract No. DE-AC05-00OR22725. 

This research used awards of computer time provided by the National Energy Research Scientific Computing Center (NERSC), a U.S. Department of Energy Office of Science User Facility located at Lawrence Berkeley National Laboratory, operated under Contract No. DE-AC02-05CH11231. 

We acknowledge EuroHPC JU for awarding this project access to LUMI-G at CSC Finland
and the Gauss Centre for Supercomputing e.V. (www.gauss-centre.eu) for funding this project by providing computing time through the John von Neumann Institute for Computing (NIC) on the GCS Supercomputer JUWELS at Jülich Supercomputing Centre (JSC).
We thank EuroHPC JU for their support within the extraordinary support program (ESP).

This work was supported by the Deutsche Forschungsgemeinschaft (DFG, German 
Research Foundation) Proj. No. 315477589-TRR 211 and the European Union under Grant Agreement No. H2020-MSCAITN-2018-813942. 
This work was partly performed in the framework of the PUNCH4NFDI consortium
supported by the Deutsche\\ Forschungsgemeinschaft (DFG, German Research Foundation) – project number 460248186 (PUNCH4NFDI).

J.K. was supported by the Deutsche
Forschungsgemeinschaft (DFG, German Research Foundation) through the
funds provided to the Sino-German Collaborative Research Center TRR110
"Symmetries and the Emergence of Structure in QCD" (DFG Project-ID 196253076 - TRR 110)

The authors gratefully acknowledge computing time provided to them on the high-performance computers Noctua2 at the NHR Center PC2. These are funded by the Federal Ministry of Education and Research and the state governments participating on the basis of the resolutions of the GWK for the national highperformance computing at universities (www.nhr-verein.de/unsere-partner).

We thank the Bielefeld HPC.NRW team for their support and M. Klappenbach for his hard work
maintaining the Bielefeld cluster. We thank H. Simma for useful 
technical discussions relating to the ILDG format.
Thanks to A. R. Gannon for the design of Figs. \ref{fig:folder} and \ref{fig:siteComm}.
Finally we extend a big thanks to G. Curell for implementing a container.

\bibliography{bibliography}

\begin{thebibliography}{10}
\expandafter\ifx\csname url\endcsname\relax
  \def\url#1{\texttt{#1}}\fi
\expandafter\ifx\csname urlprefix\endcsname\relax\def\urlprefix{URL }\fi
\expandafter\ifx\csname href\endcsname\relax
  \def\href#1#2{#2} \def\path#1{#1}\fi

\bibitem{MILC:2009mpl}
A.~Bazavov, et~al., {Nonperturbative QCD Simulations with 2+1 Flavors of
  Improved Staggered Quarks}, Rev. Mod. Phys. 82 (2010) 1349--1417.
\newblock \href {http://arxiv.org/abs/0903.3598} {\path{arXiv:0903.3598}},
  \href {http://dx.doi.org/10.1103/RevModPhys.82.1349}
  {\path{doi:10.1103/RevModPhys.82.1349}}.

\bibitem{Wilson:1974sk}
K.~G. Wilson, {Confinement of Quarks}, Phys. Rev. D 10 (1974) 2445--2459.
\newblock \href {http://dx.doi.org/10.1103/PhysRevD.10.2445}
  {\path{doi:10.1103/PhysRevD.10.2445}}.

\bibitem{Kogut:1974ag}
J.~B. Kogut, L.~Susskind, {Hamiltonian Formulation of Wilson's Lattice Gauge
  Theories}, Phys. Rev. D 11 (1975) 395--408.
\newblock \href {http://dx.doi.org/10.1103/PhysRevD.11.395}
  {\path{doi:10.1103/PhysRevD.11.395}}.

\bibitem{follana_highly_2007}
{E. Follana et al. [HPQCD, UKQCD collaboration]}, {Highly improved staggered
  quarks on the lattice, with applications to charm physics}, Phys. Rev. D 75
  (2007) 054502.
\newblock \href {http://arxiv.org/abs/hep-lat/0610092}
  {\path{arXiv:hep-lat/0610092}}, \href
  {http://dx.doi.org/10.1103/PhysRevD.75.054502}
  {\path{doi:10.1103/PhysRevD.75.054502}}.

\bibitem{Bazavov:2011nk}
A.~Bazavov, et~al., {The chiral and deconfinement aspects of the QCD
  transition}, Phys. Rev. D 85 (2012) 054503.
\newblock \href {http://arxiv.org/abs/1111.1710} {\path{arXiv:1111.1710}},
  \href {http://dx.doi.org/10.1103/PhysRevD.85.054503}
  {\path{doi:10.1103/PhysRevD.85.054503}}.

\bibitem{orginos_innovations_2006}
K.~Orginos,
  \href{https://iopscience.iop.org/article/10.1088/1742-6596/46/1/018}{Innovations
  in lattice {QCD} algorithms}, J. Phys.: Conf. Ser. 46 (2006) 132--141.
\newblock \href {http://dx.doi.org/10.1088/1742-6596/46/1/018}
  {\path{doi:10.1088/1742-6596/46/1/018}}.
\newline\urlprefix\url{https://iopscience.iop.org/article/10.1088/1742-6596/46/1/018}

\bibitem{Kronfeld:2007ek}
A.~S. Kronfeld, {Lattice Gauge Theory with Staggered Fermions: How, Where, and
  Why (Not)}, PoS LATTICE2007 (2007) 016.
\newblock \href {http://arxiv.org/abs/0711.0699} {\path{arXiv:0711.0699}},
  \href {http://dx.doi.org/10.22323/1.042.0016}
  {\path{doi:10.22323/1.042.0016}}.

\bibitem{Bernard:2007ma}
C.~Bernard, M.~Golterman, Y.~Shamir, {Effective field theories for QCD with
  rooted staggered fermions}, Phys. Rev. D 77 (2008) 074505.
\newblock \href {http://arxiv.org/abs/0712.2560} {\path{arXiv:0712.2560}},
  \href {http://dx.doi.org/10.1103/PhysRevD.77.074505}
  {\path{doi:10.1103/PhysRevD.77.074505}}.

\bibitem{Ding:2015ona}
H.-T. Ding, F.~Karsch, S.~Mukherjee, {Thermodynamics of strong-interaction
  matter from Lattice QCD}, Int. J. Mod. Phys. E 24~(10) (2015) 1530007.
\newblock \href {http://arxiv.org/abs/1504.05274} {\path{arXiv:1504.05274}},
  \href {http://dx.doi.org/10.1142/S0218301315300076}
  {\path{doi:10.1142/S0218301315300076}}.

\bibitem{Schmidt:2017bjt}
C.~Schmidt, S.~Sharma, {The phase structure of QCD}, J. Phys. G 44~(10) (2017)
  104002.
\newblock \href {http://arxiv.org/abs/1701.04707} {\path{arXiv:1701.04707}},
  \href {http://dx.doi.org/10.1088/1361-6471/aa824a}
  {\path{doi:10.1088/1361-6471/aa824a}}.

\bibitem{Guenther:2020jwe}
J.~N. Guenther, {Overview of the QCD phase diagram: Recent progress from the
  lattice}, Eur. Phys. J. A 57~(4) (2021) 136.
\newblock \href {http://arxiv.org/abs/2010.15503} {\path{arXiv:2010.15503}},
  \href {http://dx.doi.org/10.1140/epja/s10050-021-00354-6}
  {\path{doi:10.1140/epja/s10050-021-00354-6}}.

\bibitem{FermilabLattice:2019ikx}
A.~Bazavov, et~al., {$B_s\to K\ell\nu$ decay from lattice QCD}, Phys. Rev. D
  100~(3) (2019) 034501.
\newblock \href {http://arxiv.org/abs/1901.02561} {\path{arXiv:1901.02561}},
  \href {http://dx.doi.org/10.1103/PhysRevD.100.034501}
  {\path{doi:10.1103/PhysRevD.100.034501}}.

\bibitem{FermilabLattice:2022smb}
C.~T.~H. Davies, et~al., {Windows on the hadronic vacuum polarisation
  contribution to the muon anomalous magnetic moment}\href
  {http://arxiv.org/abs/2207.04765} {\path{arXiv:2207.04765}}.

\bibitem{Fan:2022kcb}
Z.~Fan, W.~Good, H.-W. Lin, {Gluon parton distribution of the nucleon from
  (2+1+1)-flavor lattice QCD in the physical-continuum limit}, Phys. Rev. D
  108~(1) (2023) 014508.
\newblock \href {http://arxiv.org/abs/2210.09985} {\path{arXiv:2210.09985}},
  \href {http://dx.doi.org/10.1103/PhysRevD.108.014508}
  {\path{doi:10.1103/PhysRevD.108.014508}}.

\bibitem{Chakraborty:2014aca}
B.~Chakraborty, C.~T.~H. Davies, B.~Galloway, P.~Knecht, J.~Koponen, G.~C.
  Donald, R.~J. Dowdall, G.~P. Lepage, C.~McNeile, {High-precision quark masses
  and QCD coupling from $n_f=4$ lattice QCD}, Phys. Rev. D 91~(5) (2015)
  054508.
\newblock \href {http://arxiv.org/abs/1408.4169} {\path{arXiv:1408.4169}},
  \href {http://dx.doi.org/10.1103/PhysRevD.91.054508}
  {\path{doi:10.1103/PhysRevD.91.054508}}.

\bibitem{MILC}
{MILC} collaboration code for lattice qcd calculations,
  \url{https://github.com/milc-qcd/milc_qcd}.

\bibitem{Clark:2009wm}
M.~A. Clark, R.~Babich, K.~Barros, R.~C. Brower, C.~Rebbi, {Solving Lattice QCD
  systems of equations using mixed precision solvers on GPUs}, Comput. Phys.
  Commun. 181 (2010) 1517--1528.
\newblock \href {http://arxiv.org/abs/0911.3191} {\path{arXiv:0911.3191}},
  \href {http://dx.doi.org/10.1016/j.cpc.2010.05.002}
  {\path{doi:10.1016/j.cpc.2010.05.002}}.

\bibitem{QUDA}
M.~A. Clark, et~al.,
  \href{https://doi.org/10.5281/zenodo.5610079}{lattice/quda: Quda v1.1.0}
  (Oct. 2021).
\newblock \href {http://dx.doi.org/10.5281/zenodo.5610079}
  {\path{doi:10.5281/zenodo.5610079}}.
\newline\urlprefix\url{https://doi.org/10.5281/zenodo.5610079}

\bibitem{Edwards:2004sx}
R.~G. Edwards, B.~Joo, {The Chroma software system for lattice QCD}, Nucl.
  Phys. B Proc. Suppl. 140 (2005) 832.
\newblock \href {http://arxiv.org/abs/hep-lat/0409003}
  {\path{arXiv:hep-lat/0409003}}, \href
  {http://dx.doi.org/10.1016/j.nuclphysbps.2004.11.254}
  {\path{doi:10.1016/j.nuclphysbps.2004.11.254}}.

\bibitem{Chroma}
The {Chroma} software system for lattice {QCD},
  \url{https://github.com/JeffersonLab/chroma/tree/master}.

\bibitem{CPS}
{USQCD: US} lattice quantum chromodynamics,
  \url{https://usqcd-software.github.io/CPS.html}.

\bibitem{Boyle:2016lbp}
P.~A. Boyle, G.~Cossu, A.~Yamaguchi, A.~Portelli, {Grid: A next generation data
  parallel C++ QCD library}, PoS LATTICE2015 (2016) 023.
\newblock \href {http://dx.doi.org/10.22323/1.251.0023}
  {\path{doi:10.22323/1.251.0023}}.

\bibitem{GRID}
Grid: Data parallel {C++} mathematical object library,
  \url{https://github.com/paboyle/Grid}.

\bibitem{openQCD}
{openQCD} simulation programs for {lattice QCD},
  \url{https://luscher.web.cern.ch/luscher/openQCD/}.

\bibitem{Mazur:2021zgi}
L.~Mazur, {Topological Aspects in Lattice QCD}, Ph.D. thesis, Bielefeld U.
  (2021).
\newblock \href {http://dx.doi.org/10.4119/unibi/2956493}
  {\path{doi:10.4119/unibi/2956493}}.

\bibitem{github}
{\texttt{SIMULATeQCD}} public code repository,
  \url{https://github.com/LatticeQCD/SIMULATeQCD}.
\newblock \href {http://dx.doi.org/10.5281/zenodo.10363968}
  {\path{doi:10.5281/zenodo.10363968}}.

\bibitem{Bollweg:2021cvl}
D.~Bollweg, L.~Altenkort, D.~A. Clarke, O.~Kaczmarek, L.~Mazur, C.~Schmidt,
  P.~Scior, H.-T. Shu, {HotQCD on multi-GPU Systems}, PoS LATTICE2021 (2022)
  196.
\newblock \href {http://arxiv.org/abs/2111.10354} {\path{arXiv:2111.10354}},
  \href {http://dx.doi.org/10.22323/1.396.0196}
  {\path{doi:10.22323/1.396.0196}}.

\bibitem{Clarke:2020htu}
D.~A. Clarke, O.~Kaczmarek, F.~Karsch, A.~Lahiri, M.~Sarkar, {Sensitivity of
  the Polyakov loop and related observables to chiral symmetry restoration},
  Phys. Rev. D 103~(1) (2021) L011501.
\newblock \href {http://arxiv.org/abs/2008.11678} {\path{arXiv:2008.11678}},
  \href {http://dx.doi.org/10.1103/PhysRevD.103.L011501}
  {\path{doi:10.1103/PhysRevD.103.L011501}}.

\bibitem{Bazavov:2020bjn}
A.~Bazavov, et~al., {Skewness, kurtosis, and the fifth and sixth order
  cumulants of net baryon-number distributions from lattice QCD confront
  high-statistics STAR data}, Phys. Rev. D 101~(7) (2020) 074502.
\newblock \href {http://arxiv.org/abs/2001.08530} {\path{arXiv:2001.08530}},
  \href {http://dx.doi.org/10.1103/PhysRevD.101.074502}
  {\path{doi:10.1103/PhysRevD.101.074502}}.

\bibitem{Bollweg:2021vqf}
D.~Bollweg, J.~Goswami, O.~Kaczmarek, F.~Karsch, S.~Mukherjee, P.~Petreczky,
  C.~Schmidt, P.~Scior, {Second order cumulants of conserved charge
  fluctuations revisited: Vanishing chemical potentials}, Phys. Rev. D 104~(7).
\newblock \href {http://arxiv.org/abs/2107.10011} {\path{arXiv:2107.10011}},
  \href {http://dx.doi.org/10.1103/PhysRevD.104.074512}
  {\path{doi:10.1103/PhysRevD.104.074512}}.

\bibitem{Bollweg:2022rps}
D.~Bollweg, J.~Goswami, O.~Kaczmarek, F.~Karsch, S.~Mukherjee, P.~Petreczky,
  C.~Schmidt, P.~Scior, {Taylor expansions and Pad\'e approximants for
  cumulants of conserved charge fluctuations at nonvanishing chemical
  potentials}, Phys. Rev. D 105~(7) (2022) 074511.
\newblock \href {http://arxiv.org/abs/2202.09184} {\path{arXiv:2202.09184}},
  \href {http://dx.doi.org/10.1103/PhysRevD.105.074511}
  {\path{doi:10.1103/PhysRevD.105.074511}}.

\bibitem{Bollweg:2022fqq}
D.~Bollweg, D.~A. Clarke, J.~Goswami, O.~Kaczmarek, F.~Karsch, S.~Mukherjee,
  P.~Petreczky, C.~Schmidt, S.~Sharma, {Equation of state and speed of sound of
  (2+1)-flavor QCD in strangeness-neutral matter at non-vanishing net
  baryon-number density}\href {http://arxiv.org/abs/2212.09043}
  {\path{arXiv:2212.09043}}.

\bibitem{Dimopoulos:2021vrk}
P.~Dimopoulos, L.~Dini, F.~Di~Renzo, J.~Goswami, G.~Nicotra, C.~Schmidt,
  S.~Singh, K.~Zambello, F.~Ziesch\'e, {Contribution to understanding the phase
  structure of strong interaction matter: Lee-Yang edge singularities from
  lattice QCD}, Phys. Rev. D 105~(3) (2022) 034513.
\newblock \href {http://arxiv.org/abs/2110.15933} {\path{arXiv:2110.15933}},
  \href {http://dx.doi.org/10.1103/PhysRevD.105.034513}
  {\path{doi:10.1103/PhysRevD.105.034513}}.

\bibitem{Cuteri:2022vwk}
F.~Cuteri, J.~Goswami, F.~Karsch, A.~Lahiri, M.~Neumann, O.~Philipsen,
  C.~Schmidt, A.~Sciarra, {Toward the chiral phase transition in the
  Roberge-Weiss plane}, Phys. Rev. D 106~(1) (2022) 014510.
\newblock \href {http://arxiv.org/abs/2205.12707} {\path{arXiv:2205.12707}},
  \href {http://dx.doi.org/10.1103/PhysRevD.106.014510}
  {\path{doi:10.1103/PhysRevD.106.014510}}.

\bibitem{Dini:2021hug}
L.~Dini, P.~Hegde, F.~Karsch, A.~Lahiri, C.~Schmidt, S.~Sharma, {Chiral phase
  transition in three-flavor QCD from lattice QCD}, Phys. Rev. D 105~(3) (2022)
  034510.
\newblock \href {http://arxiv.org/abs/2111.12599} {\path{arXiv:2111.12599}},
  \href {http://dx.doi.org/10.1103/PhysRevD.105.034510}
  {\path{doi:10.1103/PhysRevD.105.034510}}.

\bibitem{altenkort_heavy_2021}
L.~Altenkort, A.~M. Eller, O.~Kaczmarek, L.~Mazur, G.~D. Moore, H.-T. Shu,
  {Heavy quark momentum diffusion from the lattice using gradient flow}, Phys.
  Rev. D 103~(1) (2021) 014511.
\newblock \href {http://arxiv.org/abs/2009.13553} {\path{arXiv:2009.13553}},
  \href {http://dx.doi.org/10.1103/PhysRevD.103.014511}
  {\path{doi:10.1103/PhysRevD.103.014511}}.

\bibitem{sphaleronrate2021}
L.~Altenkort, A.~M. Eller, O.~Kaczmarek, L.~Mazur, G.~D. Moore, H.-T. Shu,
  \href{https://link.aps.org/doi/10.1103/PhysRevD.103.114513}{Sphaleron rate
  from euclidean lattice correlators: An exploration}, Phys. Rev. D 103 (2021)
  114513.
\newblock \href {http://dx.doi.org/10.1103/PhysRevD.103.114513}
  {\path{doi:10.1103/PhysRevD.103.114513}}.
\newline\urlprefix\url{https://link.aps.org/doi/10.1103/PhysRevD.103.114513}

\bibitem{viscosity2022}
L.~Altenkort, A.~M. Eller, A.~Francis, O.~Kaczmarek, L.~Mazur, G.~D. Moore,
  H.-T. Shu, \href{https://arxiv.org/abs/2211.08230}{Viscosity of pure-glue qcd
  from the lattice}.
\newblock \href {http://dx.doi.org/10.48550/ARXIV.2211.08230}
  {\path{doi:10.48550/ARXIV.2211.08230}}.
\newline\urlprefix\url{https://arxiv.org/abs/2211.08230}

\bibitem{Altenkort:2021jbk}
L.~Altenkort, A.~M. Eller, O.~Kaczmarek, L.~Mazur, G.~D. Moore, H.~T. Shu,
  {Lattice QCD noise reduction for bosonic correlators through blocking}, Phys.
  Rev. D 105 (2022) 094505.
\newblock \href {http://arxiv.org/abs/2112.02282} {\path{arXiv:2112.02282}},
  \href {http://dx.doi.org/10.1103/PhysRevD.105.094505}
  {\path{doi:10.1103/PhysRevD.105.094505}}.

\bibitem{Altenkort:2023oms}
L.~Altenkort, O.~Kaczmarek, R.~Larsen, S.~Mukherjee, P.~Petreczky, H.-T. Shu,
  S.~Stendebach, {Heavy Quark Diffusion from 2+1 Flavor Lattice QCD}\href
  {http://arxiv.org/abs/2302.08501} {\path{arXiv:2302.08501}}.

\bibitem{Clarke:2019tzf}
D.~A. Clarke, O.~Kaczmarek, F.~Karsch, A.~Lahiri, {Polyakov Loop Susceptibility
  and Correlators in the Chiral Limit}, PoS LATTICE2019 (2020) 194.
\newblock \href {http://arxiv.org/abs/1911.07668} {\path{arXiv:1911.07668}},
  \href {http://dx.doi.org/10.22323/1.363.0194}
  {\path{doi:10.22323/1.363.0194}}.

\bibitem{GaurangParkar:2022aft}
G.~Parkar, D.~Bala, O.~Kaczmarek, R.~Larsen, S.~Mukherjee, P.~Petreczky,
  A.~Rothkopf, J.~H. Weber, {Static quark anti-quark interactions at non-zero
  temperature from lattice QCD}, EPJ Web Conf. 274 (2022) 04006.
\newblock \href {http://arxiv.org/abs/2211.12937} {\path{arXiv:2211.12937}},
  \href {http://dx.doi.org/10.1051/epjconf/202227404006}
  {\path{doi:10.1051/epjconf/202227404006}}.

\bibitem{GaurangParkar:2022aoa}
G.~Parkar, O.~Kaczmarek, R.~Larsen, S.~Mukherjee, P.~Petreczky, A.~Rothkopf,
  J.~H. Weber, {Complex potential at T \ensuremath{>} 0 from fine lattices},
  PoS LATTICE2022 (2023) 188.
\newblock \href {http://dx.doi.org/10.22323/1.430.0188}
  {\path{doi:10.22323/1.430.0188}}.

\bibitem{metropolis_equation_1953}
N.~Metropolis, A.~W. Rosenbluth, M.~N. Rosenbluth, A.~H. Teller, E.~Teller,
  \href{http://aip.scitation.org/doi/10.1063/1.1699114}{Equation of {State}
  {Calculations} by {Fast} {Computing} {Machines}}, J. Chem. Phys. 21~(6)
  (1953) 1087--1092.
\newblock \href {http://dx.doi.org/10.1063/1.1699114}
  {\path{doi:10.1063/1.1699114}}.
\newline\urlprefix\url{http://aip.scitation.org/doi/10.1063/1.1699114}

\bibitem{press_numerical_1992}
W.~H. Press, S.~A. Teukolsky, W.~T. Vetterling, B.~P. Flannery, Numerical
  recipes: the art of scientific computing, 2nd Edition, 1992.

\bibitem{lecuyer_maximally_1996}
P.~L’Ecuyer,
  \href{https://www.ams.org/mcom/1996-65-213/S0025-5718-96-00696-5/}{Maximally
  equidistributed combined {Tausworthe} generators}, Math. Comp. 65~(213)
  (1996) 203--213.
\newblock \href {http://dx.doi.org/10.1090/S0025-5718-96-00696-5}
  {\path{doi:10.1090/S0025-5718-96-00696-5}}.
\newline\urlprefix\url{https://www.ams.org/mcom/1996-65-213/S0025-5718-96-00696-5/}

\bibitem{Hasenfratz:1983ba}
P.~Hasenfratz, F.~Karsch, {Chemical Potential on the Lattice}, Phys. Lett. B
  125 (1983) 308--310.
\newblock \href {http://dx.doi.org/10.1016/0370-2693(83)91290-X}
  {\path{doi:10.1016/0370-2693(83)91290-X}}.

\bibitem{cabibbo_new_1982}
N.~Cabibbo, E.~Marinari, A new method for updating {SU}({N}) matrices in
  computer simulations of gauge theories, Phys. Lett. B 119~(4-6) (1982)
  387--390.
\newblock \href {http://dx.doi.org/10.1016/0370-2693(82)90696-7}
  {\path{doi:10.1016/0370-2693(82)90696-7}}.

\bibitem{fabricius_heat_1984}
K.~Fabricius, O.~Haan,
  \href{http://linkinghub.elsevier.com/retrieve/pii/0370269384915028}{Heat bath
  method for the twisted {Eguchi}-{Kawai} model}, Phys. Lett. B 143~(4-6)
  (1984) 459--462.
\newblock \href {http://dx.doi.org/10.1016/0370-2693(84)91502-8}
  {\path{doi:10.1016/0370-2693(84)91502-8}}.
\newline\urlprefix\url{http://linkinghub.elsevier.com/retrieve/pii/0370269384915028}

\bibitem{kennedy_improved_1985}
A.~D. Kennedy, B.~J. Pendleton, Improved heatbath method for {Monte} {Carlo}
  calculations in lattice gauge theories, Phys. Lett. B 156~(5-6) (1985)
  393--399.
\newblock \href
  {http://dx.doi.org/https://doi.org/10.1016/0370-2693(85)91632-6}
  {\path{doi:https://doi.org/10.1016/0370-2693(85)91632-6}}.

\bibitem{adler_over-relaxation_1981}
S.~L. Adler, Over-relaxation method for the {Monte} {Carlo} evaluation of the
  partition function for multiquadratic actions, Phys. Rev. D 23~(12) (1981)
  2901--2904.
\newblock \href {http://dx.doi.org/10.1103/PhysRevD.23.2901}
  {\path{doi:10.1103/PhysRevD.23.2901}}.

\bibitem{creutz_overrelaxation_1987}
M.~Creutz, Overrelaxation and {Monte} {Carlo} simulation, Phys. Rev. D 36~(2)
  (1987) 515--519.
\newblock \href {http://dx.doi.org/10.1103/PhysRevD.36.515}
  {\path{doi:10.1103/PhysRevD.36.515}}.

\bibitem{Follana:2006rc}
E.~Follana, Q.~Mason, C.~Davies, K.~Hornbostel, G.~P. Lepage, J.~Shigemitsu,
  H.~Trottier, K.~Wong, {Highly improved staggered quarks on the lattice, with
  applications to charm physics}, Phys. Rev. D 75 (2007) 054502.
\newblock \href {http://arxiv.org/abs/hep-lat/0610092}
  {\path{arXiv:hep-lat/0610092}}, \href
  {http://dx.doi.org/10.1103/PhysRevD.75.054502}
  {\path{doi:10.1103/PhysRevD.75.054502}}.

\bibitem{Kennedy:1998cu}
A.~D. Kennedy, I.~Horvath, S.~Sint, {A New exact method for dynamical fermion
  computations with nonlocal actions}, Nucl. Phys. B Proc. Suppl. 73 (1999)
  834--836.
\newblock \href {http://arxiv.org/abs/hep-lat/9809092}
  {\path{arXiv:hep-lat/9809092}}, \href
  {http://dx.doi.org/10.1016/S0920-5632(99)85217-7}
  {\path{doi:10.1016/S0920-5632(99)85217-7}}.

\bibitem{clark_rhmc_2004}
M.~A. Clark, A.~D. Kennedy, {The RHMC algorithm for two flavors of dynamical
  staggered fermions}, Nucl. Phys. B Proc. Suppl. 129 (2004) 850--852.
\newblock \href {http://arxiv.org/abs/hep-lat/0309084}
  {\path{arXiv:hep-lat/0309084}}, \href
  {http://dx.doi.org/10.1016/S0920-5632(03)02732-4}
  {\path{doi:10.1016/S0920-5632(03)02732-4}}.

\bibitem{Mukherjee:2014pye}
S.~Mukherjee, O.~Kaczmarek, C.~Schmidt, P.~Steinbrecher, M.~Wagner, {HISQ
  inverter on Intel$\scriptsize{^{\circledR}}$ Xeon Phi$\scriptsize{^{TM}}$ and
  NVIDIA$\scriptsize{^{\circledR}}$ GPUs}, PoS LATTICE2014 (2015) 044.
\newblock \href {http://arxiv.org/abs/1409.1510} {\path{arXiv:1409.1510}},
  \href {http://dx.doi.org/10.22323/1.214.0044}
  {\path{doi:10.22323/1.214.0044}}.

\bibitem{Jegerlehner:1996pm}
B.~Jegerlehner, {Krylov space solvers for shifted linear systems}\href
  {http://arxiv.org/abs/hep-lat/9612014} {\path{arXiv:hep-lat/9612014}}.

\bibitem{Sexton:1992nu}
J.~C. Sexton, D.~H. Weingarten, {Hamiltonian evolution for the hybrid Monte
  Carlo algorithm}, Nucl. Phys. B 380 (1992) 665--677.
\newblock \href {http://dx.doi.org/10.1016/0550-3213(92)90263-B}
  {\path{doi:10.1016/0550-3213(92)90263-B}}.

\bibitem{Hasenbusch:2001ne}
M.~Hasenbusch, {Speeding up the hybrid Monte Carlo algorithm for dynamical
  fermions}, Phys. Lett. B 519 (2001) 177--182.
\newblock \href {http://arxiv.org/abs/hep-lat/0107019}
  {\path{arXiv:hep-lat/0107019}}, \href
  {http://dx.doi.org/10.1016/S0370-2693(01)01102-9}
  {\path{doi:10.1016/S0370-2693(01)01102-9}}.

\bibitem{OMELYAN2003272}
I.~Omelyan, I.~Mryglod, R.~Folk,
  \href{https://www.sciencedirect.com/science/article/pii/S0010465502007543}{Symplectic
  analytically integrable decomposition algorithms: classification, derivation,
  and application to molecular dynamics, quantum and celestial mechanics
  simulations}, Computer Physics Communications 151~(3) (2003) 272--314.
\newblock \href
  {http://dx.doi.org/https://doi.org/10.1016/S0010-4655(02)00754-3}
  {\path{doi:https://doi.org/10.1016/S0010-4655(02)00754-3}}.
\newline\urlprefix\url{https://www.sciencedirect.com/science/article/pii/S0010465502007543}

\bibitem{MILC:2010pul}
A.~Bazavov, et~al., {Scaling studies of QCD with the dynamical HISQ action},
  Phys. Rev. D 82 (2010) 074501.
\newblock \href {http://arxiv.org/abs/1004.0342} {\path{arXiv:1004.0342}},
  \href {http://dx.doi.org/10.1103/PhysRevD.82.074501}
  {\path{doi:10.1103/PhysRevD.82.074501}}.

\bibitem{bilson-thompson_highly_2003}
S.~O. Bilson-Thompson, D.~B. Leinweber, A.~G. Williams,
  \href{https://linkinghub.elsevier.com/retrieve/pii/S0003491603000095}{Highly
  improved lattice field-strength tensor}, Annals of Physics 304~(1) (2003)
  1--21.
\newblock \href {http://dx.doi.org/10.1016/S0003-4916(03)00009-5}
  {\path{doi:10.1016/S0003-4916(03)00009-5}}.
\newline\urlprefix\url{https://linkinghub.elsevier.com/retrieve/pii/S0003491603000095}

\bibitem{mandula_efficient_1990}
J.~E. Mandula, M.~Ogilvie, Efficient gauge fixing via overrelaxation, Phys.
  Lett. B 248~(1-2) (1990) 156--158.
\newblock \href
  {http://dx.doi.org/https://doi.org/10.1016/0370-2693(90)90031-Z}
  {\path{doi:https://doi.org/10.1016/0370-2693(90)90031-Z}}.

\bibitem{luscher_properties_2010}
M.~L\"uscher, {Properties and uses of the Wilson flow in lattice QCD}, JHEP 08
  (2010) 071, [Erratum: JHEP 03, 092 (2014)].
\newblock \href {http://arxiv.org/abs/1006.4518} {\path{arXiv:1006.4518}},
  \href {http://dx.doi.org/10.1007/JHEP08(2010)071}
  {\path{doi:10.1007/JHEP08(2010)071}}.

\bibitem{ramos_symanzik_2016}
A.~Ramos, S.~Sint, {Symanzik improvement of the gradient flow in lattice gauge
  theories}, Eur. Phys. J. C 76~(1) (2016) 15.
\newblock \href {http://arxiv.org/abs/1508.05552} {\path{arXiv:1508.05552}},
  \href {http://dx.doi.org/10.1140/epjc/s10052-015-3831-9}
  {\path{doi:10.1140/epjc/s10052-015-3831-9}}.

\bibitem{munthe-kaas_runge-kutta_1998}
H.~Munthe-Kaas,
  \href{http://link.springer.com/10.1007/BF02510919}{Runge-{Kutta} methods on
  {Lie} groups}, Bit Numer Math 38~(1) (1998) 92--111.
\newblock \href {http://dx.doi.org/10.1007/BF02510919}
  {\path{doi:10.1007/BF02510919}}.
\newline\urlprefix\url{http://link.springer.com/10.1007/BF02510919}

\bibitem{celledoni_commutator-free_2003}
E.~Celledoni, A.~Marthinsen, B.~Owren,
  \href{https://linkinghub.elsevier.com/retrieve/pii/S0167739X02001619}{Commutator-free
  {Lie} group methods}, Future Generation Computer Systems 19~(3) (2003)
  341--352.
\newblock \href {http://dx.doi.org/10.1016/S0167-739X(02)00161-9}
  {\path{doi:10.1016/S0167-739X(02)00161-9}}.
\newline\urlprefix\url{https://linkinghub.elsevier.com/retrieve/pii/S0167739X02001619}

\bibitem{Hasenfratz:2001hp}
A.~Hasenfratz, F.~Knechtli, {Flavor symmetry and the static potential with
  hypercubic blocking}, Phys. Rev. D 64 (2001) 034504.
\newblock \href {http://arxiv.org/abs/hep-lat/0103029}
  {\path{arXiv:hep-lat/0103029}}, \href
  {http://dx.doi.org/10.1103/PhysRevD.64.034504}
  {\path{doi:10.1103/PhysRevD.64.034504}}.

\bibitem{luscher_locality_2001}
M.~Luscher, P.~Weisz, {Locality and exponential error reduction in numerical
  lattice gauge theory}, JHEP 09 (2001) 010.
\newblock \href {http://arxiv.org/abs/hep-lat/0108014}
  {\path{arXiv:hep-lat/0108014}}, \href
  {http://dx.doi.org/10.1088/1126-6708/2001/09/010}
  {\path{doi:10.1088/1126-6708/2001/09/010}}.

\bibitem{meyer_locality_2003}
H.~B. Meyer, {Locality and statistical error reduction on correlation
  functions}, JHEP 01 (2003) 048.
\newblock \href {http://arxiv.org/abs/hep-lat/0209145}
  {\path{arXiv:hep-lat/0209145}}, \href
  {http://dx.doi.org/10.1088/1126-6708/2003/01/048}
  {\path{doi:10.1088/1126-6708/2003/01/048}}.

\bibitem{Veldhuizen:1995}
T.~Veldhuizen,
  \href{https://web.archive.org/web/20050210090012/http://osl.iu.edu/~tveldhui/papers/Expression-Templates/exprtmpl.html}{Expression
  templates}, C++ Report 7~(5) (1995) 26--31.
\newline\urlprefix\url{https://web.archive.org/web/20050210090012/http://osl.iu.edu/~tveldhui/papers/Expression-Templates/exprtmpl.html}

\bibitem{beckett_building_2011}
M.~G. Beckett, P.~Coddington, B.~Joó, C.~M. Maynard, D.~Pleiter, O.~Tatebe,
  T.~Yoshie,
  \href{https://linkinghub.elsevier.com/retrieve/pii/S0010465511000476}{Building
  the {International} {Lattice} {Data} {Grid}}, Computer Physics Communications
  182~(6) (2011) 1208--1214.
\newblock \href {http://dx.doi.org/10.1016/j.cpc.2011.01.027}
  {\path{doi:10.1016/j.cpc.2011.01.027}}.
\newline\urlprefix\url{https://linkinghub.elsevier.com/retrieve/pii/S0010465511000476}

\bibitem{maynard_qcdml_2005}
C.~Maynard, D.~Pleiter,
  \href{https://linkinghub.elsevier.com/retrieve/pii/S0920563204006814}{{QCDml}:
  {First} milestones for building an {International} {Lattice} {Data} {Grid}},
  Nuclear Physics B - Proceedings Supplements 140 (2005) 213--221.
\newblock \href {http://dx.doi.org/10.1016/j.nuclphysbps.2004.11.116}
  {\path{doi:10.1016/j.nuclphysbps.2004.11.116}}.
\newline\urlprefix\url{https://linkinghub.elsevier.com/retrieve/pii/S0920563204006814}

\end{thebibliography}

\end{document}